\documentclass[namedreferences]{solarphysics}

\usepackage[optionalrh]{spr-sola-addons} % For Solar Physics
\usepackage{natbib}          % natbib
\usepackage{graphicx,epstopdf}        % For eps figures, newer & more powerfull
\usepackage{amssymb}        % useful mathematical symbols
\usepackage{color}           % For color text: \color command
\usepackage{array}
\newcolumntype{C}[1]{>{\centering\arraybackslash}p{#1}}
\usepackage{url}

\usepackage{color}

\newcommand{\aap}{    {\it Astron. Astrophys.}}

\newcommand{\apj}{    {\it Astrophys. J.}}

\newcommand{\ssr}{    {\it Space Sci. Rev.}}
\chardef\us=`\_

\begin{document}

\begin{article}
\begin{opening}

\title{Self-consistent Modelling of Gamma-Ray Spectra from Solar Flares with the Monte Carlo Simulation Package FLUKA\\ {\it Solar Physics}}

\author[addressref={aff1},email={danesaracol@gmail.com}]{\inits{D.S.}\fnm{Daneele S.}~\lnm{Tusnski}}%\sep
\author[addressref=aff1,corref,email={sergio.szpigel@mackenzie.br}]{\inits{S.}\fnm{Sergio}~\lnm{Szpigel}\orcid{0000-0003-2529-2225}}%\sep
\author[addressref={aff1,aff2},email={guigue@craam.mackenzie.br}]{\inits{C.G.}\fnm{Carlos Guillermo}~\lnm{Gim\'enez de Castro}\orcid{0000-0002-8979-3582}}%\sep
\author[addressref=aff3,email={alexander.mackinnon@glasgow.ac.uk}]{\inits{A.L.}\fnm{Alexander L.}~\lnm{MacKinnon}\orcid{0000-0002-3558-4806}}%\sep
\author[addressref={aff1,aff3},email={paulo.simoes@glasgow.ac.uk}]{\inits{P.J.A.}\fnm{Paulo Jos\'e A.}~\lnm{Sim\~oes}\orcid{0000-0002-4819-1884}}
\address[id=aff1]{Centro de R\'adio-Astronomia e Astrof\'isica Mackenzie (CRAAM), Escola de Engenharia, Universidade Presbiteriana Mackenzie, S\~ao Paulo, Brazil}
\address[id=aff2]{Instituto de Astronom\'{\i}a y F\'{\i}sica del Espacio, CONICET, Buenos Aires, Argentina.}
\address[id=aff3]{School of Physics and Astronomy, University of Glasgow, Glasgow, UK}

\runningauthor{D. S. Tusnski {\it et al.}}
\runningtitle{Self-consistent Modelling of Gamma-Ray Spectra}

\begin{abstract}

    We use the Monte Carlo particle physics code FLUKA ({\it Fluktuierende Kaskade}) to calculate $\gamma$-ray spectra expected from solar flare energetic ion distributions. The FLUKA code includes robust physics-based models for electromagnetic, hadronic and nuclear interactions, sufficiently detailed for it to be a useful tool for calculating nuclear de-excitation, positron annihilation and neutron capture line fluxes and shapes, as well as $\approx \; {\rm GeV}$ continuum radiation from pion decay products. We show nuclear de-excitation $\gamma$-ray line model spectra from a range of assumed primary accelerated ion distributions and find them to be in good agreement with those found using the code of \cite{Murphy2009}. We also show full $\gamma$-ray model spectra which exhibit all the typical structures of $\gamma$-ray spectra observed in solar flares. From these model spectra we build templates which are incorporated into the software package {\it Objective Spectral Executive} (OSPEX) and used to fit the combined {\it Fermi Gamma-ray Burst Monitor} (GBM)/{\it Large Area Telescope} (LAT) spectrum of the 2010 June 12 solar flare, providing a statistically acceptable result. To the best of our knowledge, the fit carried out with the FLUKA templates for the full $\gamma$-ray spectrum can be regarded as the first attempt to use a single code to implement a self-consistent treatment of the several spectral components in the photon energy range from $\approx 100$s ${\rm keV}$ to $\approx 100$s ${\rm MeV}$.

\end{abstract}
\end{opening}

%-------------------------------------------------

\section{Introduction}
     \label{S-Introduction}

Particle acceleration plays a key role in solar flares. Accelerated particles may constitute a substantial fraction of all the energy released in a flare \cite[]{Emslie2004,Emslie2005}. Their radiative signatures are characteristic of the flare impulsive phase and clearly key to the energy release process \cite[]{Kane1974}. Moreover, empirical knowledge of flares serves as ``ground truth'' for other astrophysical energy release events in more remote settings. Energetic particles are detected {\it in situ} in interplanetary space but we are reliant on electromagnetic radiation to learn about particle acceleration at the flare site and it is important that we develop reliable methods to interpret spectral, spatial and timing information.

Observations of (electron bremsstrahlung) flare hard X-rays with the {\it Ramaty High-Energy Spectroscopic Imager} (RHESSI) satellite have yielded a refined characterisation of flare electron acceleration and its relationship to other flare processes \cite[]{Holman2011,Kontar2011}. The picture for flare ions remains less complete, partly because flares producing detectable $\gamma$-rays are much rarer but also because of the greater complexity of the nuclear radiation mechanisms. Ion acceleration in flares is studied {\it via} a variety of signatures at $\gamma$-ray wavelengths, consequences of nuclear reactions of primary accelerated ions with ambient nuclei \cite[]{Share2006,Murphy2007,Vilmer2011}: nuclear de-excitation lines in the $0.4-7 \;{\rm MeV}$ range, excited by ions of $1-100\;{\rm MeV/nucleon}$ \cite[]{Ramaty1979,Kozlovsky2002}; $2.223\;{\rm MeV}$ neutron capture and $511\;{\rm keV}$ positron annihilation lines, each resulting from ions spanning a wide energy range, from a few to $100$s of ${\rm MeV/nucleon}$ \cite[]{Lockwood1997,Hua2002,Kozlovsky1987,Kozlovsky2002,Murphy2007}; continuum radiation in the $\gtrsim 0.1\;{\rm MeV}$ range, resulting from pion decay products of $\gtrsim 0.2 - 0.3 \;{\rm GeV/nucleon}$ ions \cite[]{Dermer1986,Murphy1987,Mandhzavidze1992,Vilmer2003}. Energetic neutrons detected in space or with ground-based instruments also give information on accelerated flare ions \cite[{\it e.g.}]{Chupp1987,Kocharov1998,Watanabe2006}.

Interpretation of observed $\gamma$-ray spectra must proceed {\it via} best-fit from among a set of templates that span a range of assumptions about the ion energy distributions and source chemical abundances, and may also involve details of magnetic geometry, {\it etc}. \cite[]{Murphy2007,Vilmer2011}. The generation and use of such templates has been developed over several decades, starting with the seminal work of  \cite{Lingenfelter1967}. A detailed study of the production of nuclear de-excitation $\gamma$-ray lines was carried out by \cite{Ramaty1979}, resulting in the development of a Monte Carlo code for calculating complete spectra of de-excitation $\gamma$-ray lines. The code includes both the ``direct'' reactions of energetic protons, $^3$He and $\alpha$-particles with ambient nuclei and the ``inverse'' reactions of energetic nuclei heavier than $\alpha$-particles with ambient H and $^4$He. This code was updated by \cite{Kozlovsky2002} with new cross section data from laboratory measurements. An improved version of the code (hereafter referred to as RMK) was further developed by \cite{Murphy2009} incorporating cross sections evaluated with the nuclear reaction code TALYS \cite[]{Koning2005}. Templates for the nuclear de-excitation $\gamma$-ray lines are available as standard fit functions in the software package {\it Objective Spectral Executive} (OSPEX) \cite[]{Freeland:1998}, which were built from calculations with the RMK code assuming a downward isotropic angular distribution of primary accelerated ions having a power-law energy distribution with spectral index of $4.0$, a coronal abundance of accelerated ions with $\alpha/{\rm proton}$ ratio of $0.22$, a coronal abundance of ambient nuclei with ${\rm He/H}$ ratio of $0.1$, and a heliocentric angle of $60^{\circ}$. Templates for the continuum radiation from pion decay are also available in OSPEX, which were built from calculations of pion decay $\gamma$-ray spectra carried out by \cite{Murphy1987} assuming an isotropic angular distribution of primary accelerated protons having a power-law energy distribution with several spectral indexes ($3.0,\; 3.5, \; 4.0, \; 4.5, \; {\rm and} \; 5.0$), and $\alpha/{\rm proton}$ ratio of $0.1$. OSPEX provides a specific standard fit function for the component due to the $511 \;{\rm keV}$ positron annihilation line and its associated positronium continuum, but there is no specific standard fit function available for the component due to the $2.223 \;{\rm MeV}$ neutron capture line. The neutron capture line component is usually included as a gaussian function. A more accurate neutron capture line component also incorporating the associated Compton-scattered continuum can be included if desired through user-supplied templates built from calculations with codes such as the neutron-production code developed by \cite{Hua2002} and the Monte Carlo N-Particle Transport Code (MCNP6) developed by \cite{Pelowitz2013} (see {\it e.g.} \cite{Murphy2018}). OSPEX also provides a standard fit function for the component due to the $\alpha-\alpha$ lines, which was built from calculations assuming a downward isotropic distribution of primary accelerated ions having a power-law energy distribution with spectral index of $3.5$ and a heliocentric angle of $60^{\circ}$.

The Monte Carlo code FLUKA ({\it Fluktuierende Kaskade}) \cite[]{Ferrari2011,Battistoni2015} is a general purpose package of integrated routines for simulation of particle transport and interactions in arbitrary materials. It is ``not a toolkit'': it aims to provide a unified description of all relevant processes, to a consistent degree of approximation, relevant to a particular situation. It has seen a wide variety of applications in areas such as high-energy experimental physics, nuclear physics and engineering, detector and telescope design, and medical physics, as well in applications particularly close to our interests here: calculating consequences of cosmic-ray interaction in the terrestrial atmosphere \cite[]{Battistoni2008,Claret2014}; interpretation of {\it Fermi Large Area Telescope} (LAT) measurements of $\gamma$-rays produced by cosmic-ray impact on the Moon \cite[]{Ackermann2016}.

Recent contributions have begun to explore FLUKA's use as a tool for interpretation of solar flare radiations at $\gamma$-ray and other wavelengths. Since FLUKA combines physics-based models for nuclear processes with transport of primary and secondary particles and photons, \cite{MacKinnon2014} and \cite{MacKinnon2016} were able to use it to study the effects of primary particle directionality on emergent spectra. They concentrated on the $\gamma$-ray continuum at $\approx$ GeV energies resulting from neutral pion decay and from bremsstrahlung of charged pion decay positrons and electrons. It was also used to calculate the distributions of charged secondaries emerging from a photospheric thick target, with a view to evaluating their synchrotron radiation at ${\rm mm}$ and shorter wavelengths \cite[]{Tuneu2017}. Here we concentrate particularly on FLUKA's capabilities for predicting de-excitation line fluxes and present, for the first time, full, self-consistent simulations of the $\gamma$-ray spectrum, including the nuclear de-excitation lines, the $2.223\;{\rm MeV}$ neutron capture line and the pion decay continuum. Our spectra also include the $511 \; {\rm keV}$ positron annihilation line from pion decay products and pair creation in the solar atmosphere by $\gamma$-ray photons but exclude the normally dominant contribution from secondary radioactive positron-emitting nuclei. FLUKA is capable of modelling positron production via radioactive nuclei but we defer a detailed study of all components of the annihilation line for future work, noting that a full treatment of the shape of the line, the relative intensities of line and three-photon continuum, {\it etc.}, need a description of solar atmospheric properties, particularly degree of ionization, that we do not attempt to implement in FLUKA (see \cite{Murphy2005}). With this qualification, the resulting spectra offer the possibility of an independent verification and validation of the work of \cite{Ramaty1979}, \cite{Kozlovsky2002}, {\it etc}. De-excitation line production in FLUKA occurs at the end of a sequence of nuclear processes that model hadron-nucleus and nucleus-nucleus interaction. Thus the fluxes and line shapes resulting from many sorts of reactions (inelastic excitation as well as spallation and from both direct and inverse reactions) are modeled in detail. FLUKA's several physics models are detailed in the Appendix, along with several references which provide their validation by comparison to laboratory measurements.

The Monte Carlo code {\it Geometry and Tracking} (GEANT4) \cite[]{Agostinelli2003} is also rooted in laboratory expertise and is employed in a similarly wide range of contexts. \cite{Tang2010} used GEANT4 in a similar spirit to our work here, concentrating on calculating the pion decay continuum radiation from postulated primary accelerated ion distributions, as well as the $511\;{\rm keV}$ positron annihilation line. They did not use it to calculate de-excitation $\gamma$-ray line spectra. Since then, the {\it Fermi Gamma-ray Burst Monitor} (GBM) and LAT instruments have obtained good $\gamma$-ray spectra which can be compared with such calculations, as we do here. It is important to emphasize that the main purpose of the paper is to present the results of a first investigation on FLUKA's potential as an effective tool for the modelling of $\gamma$-ray spectra from solar flares. A thorough analysis to validate FLUKA for such modelling would require calculations which are beyond the scope of the paper.

In Section~\ref{s-model} we describe in detail the geometrical model we used to simulate $\gamma$-rays from flare ions with FLUKA. In Section~\ref{S-simulations} we show illustrative nuclear de-excitation $\gamma$-ray line spectra and full $\gamma$-ray spectra obtained from this model. In Section~\ref{S-fitting} we show a detailed fit to {\it Fermi} data. Since FLUKA provides an integrated description of all relevant processes we can attempt to provide a single fit to a full $\gamma$-ray spectrum stretching from $< 1 \;{\rm MeV}$ to ${\rm GeV}$ photon energies, all emerging from the same physical situation. To this end we combine {\it Fermi} GBM and LAT data to give a single spectrum including both lines and pion decay continuum ({\it cf.} \cite{Vilmer2003}). Section~\ref{S-summary} gives a brief summary of our approach and results and discusses their usefulness and implications.

%%%%%%%%%%%%%%%%%%%%%%%%%%%%%%%%%%%%%%%%%%%%%%%%%%%%%%%%%%%%%%%%%

\section{Solar Flare Model}
\label{s-model}

In our simulations of nuclear processes in solar flares using the FLUKA package, we consider a model in which beams of primary accelerated ions are injected into a target with characteristics similar to those of the ambient solar atmosphere. FLUKA's combinatorial geometry package \cite[]{Ferrari2011}, which is an improved version of the package developed for the neutron and $\gamma$-ray transport code MORSE \cite[]{Emmett1975}, allows one to set up simulations with arbitrarily complex geometries. A huge variety of geometries can be built through combinations of solid bodies and surfaces obtained by boolean operations. We adopt a simple plane-parallel geometry for the vertical structure of the ambient solar atmosphere, since in general the dimensions of the solar flare emission region are much smaller than the solar radius\footnote{The assumption of plane-parallel geometry is adequate over most of the disk. For flares very near the limb it over-estimates the column density met by escaping photons and must be replaced by a spherical geometry, straightforwardly implemented in FLUKA if necessary.}. Using the combinatorial geometry tools provided by FLUKA, we build a cubic box centered at the origin of a cartesian coordinate system (${\rm O}x,{\rm O}y,{\rm O}z$) with edges of length $L=2\times 10^9 \;{\rm cm}$ and faces perpendicular to the coordinate axes. The $z$-coordinate corresponds to the vertical depth in the ambient solar atmosphere. A $xy$-plane at $z=0$ divides the cubic box into two half-spaces. The half-space at $z<0$ represents the coronal region and, for simplicity, is just filled with vacuum (such that the particles are transported but no longer interact). The half-space at $z>0$ represents the chromospheric/photospheric region and is filled with a dense, neutral material for which we assume a typical solar atmosphere composition with the abundances of $^4$He, C, N, O, Ne, Mg, Al, Si, S, Ca and Fe nuclei relative to H given by \cite{Asplund2009}, as indicated in Table \ref{tab:compAtSolarModIII}. Our assumption of a neutral medium is appropriate to the photosphere and consistent with previous treatments of $\gamma$-ray line production ({\it e.g.} \cite{Murphy2007}). The beams of primary accelerated ions are injected into the chromospheric/photospheric region from a point ($0,0,z_0$) located in the coronal region at a distance $z_0$ above and very close to the $xy$-plane at $z=0$, avoiding any artefacts or errors that might result if the source were placed exactly on the boundary between two regions.
\begin{table}[t]
\caption{Compositions for the ambient atmosphere and the accelerated ions.}
\begin{tabular}{cccc}
 \hline
 Element  & Ambient atmosphere & Acc. ions (photospheric) & Acc. ions (impulsive flare)\\
  \hline
    H        &  $1.0$ & $1.0$ & $1.0$ \\
    He  &  $8.50 \times 10^{-2}$ &  $0.1; 0.5$ &$0.1; 0.5$ \\
    C   &  $2.69 \times 10^{-4}$ &  $2.69 \times 10^{-4}$   &$4.65 \times 10^{-3}$   \\
    N   &  $6.76 \times 10^{-5}$ &  $6.76 \times 10^{-5}$   &$1.24 \times 10^{-3}$   \\
    O   &  $4.90 \times 10^{-4}$ &  $4.90 \times 10^{-4}$   &$1.00 \times 10^{-2}$   \\
    Ne  &  $8.51 \times 10^{-5}$ &  $8.51 \times 10^{-5}$   &$4.55 \times 10^{-3}$   \\
    Mg  &  $3.98 \times 10^{-5}$ &  $3.98 \times 10^{-5}$   &$5.89 \times 10^{-3}$   \\
    Al  &  $2.82 \times 10^{-6}$ &  $2.82 \times 10^{-6}$   &$1.57 \times 10^{-4}$   \\
    Si  &  $3.24 \times 10^{-5}$ &  $3.24 \times 10^{-5}$   &$4.55 \times 10^{-3}$   \\
    S   &  $1.32 \times 10^{-5}$ &  $1.32 \times 10^{-5}$   &$9.56 \times 10^{-4}$   \\
    Ca  &  $2.19 \times 10^{-6}$ &  $2.19 \times 10^{-6}$   &$1.06 \times 10^{-4}$   \\
    Fe  &  $3.16 \times 10^{-5}$ &  $3.16 \times 10^{-5}$   &$1.34 \times 10^{-2}$   \\
\hline
\end{tabular}
\label{tab:compAtSolarModIII}
\end{table}

The half-space at $z>0$ is further divided into $52$ layers with a vertical density profile given by the semi-empirical VAL-C model of the chromosphere \cite[]{VAL1981} plus an additional layer, corresponding to the photosphere, with a density of $3.19\times 10^{-7}\; {\rm g/cm^3}$. The densities of the chromospheric layers given in the VAL-C model range from $2.35\times10^{-15}\; {\rm g/cm^3}$ at the outermost layer to $3.19\times 10^{-7}\; {\rm g/cm^3}$ at the innermost one. However, materials with densities $< 10^{-10}\; {\rm g/cm^3}$ are treated as vacuum by FLUKA. In order to circumvent this feature, we assign a density of $2 \times 10^{-10}\; {\rm g/cm^3}$ to all layers with densities $< 10^{-10}\; {\rm g/cm^3}$ and rescale their thickness while keeping the values for the column-densities as given in the VAL-C model. The total column-density obtained by summing up the contributions of all $53$ layers is $292\; {\rm g/cm^2}$, corresponding to the stopping-depth (or range) for a proton with energy of $\approx 1.6 \;{\rm GeV}$ in a hydrogen target evaluated in the continuous slowing down approximation (CSDA) \cite[]{Berger1999}. In this way, the half-space at $z>0$ fulfils the thick-target condition for primary accelerated ions with kinetic energies up to $1.6 \;{\rm GeV/nucleon}$.

It is important to note that the modification we have implemented in the density structure (with respect to the original VAL-C model) does not introduce significant changes in the results of our present simulations, since the thick target photon yields produced by all relevant processes, with the exception of particle decay processes, depend only on the total column-density. The decay processes depend on time, and thus on the distance travelled by the decaying particles within the target. However, the decay processes which are most relevant to our simulations are the decay of neutral pions ($\pi^0$), charged pions ($\pi^{\pm}$) and muons ($\mu^{\pm}$), whose lifetimes are respectively $\tau_{\pi^0}=8.5\times 10^{-17}\; {\rm s}$, $\tau_{\pi^{\pm}}=2.6\times 10^{-8}\; {\rm s}$ and  $\tau_{\mu^{\pm}}=2.2\times 10^{-6} \; {\rm s}$. A muon or pion with speed very close to $c$ thus travels $< 1$ km in a rest-frame lifetime. Primary protons with energies up to $10 \; {\rm GeV}$ can produce muons with Lorentz gamma factor up to $\approx 70$ (using results of \cite{Dermer1987}) but this still means they travel $\lesssim 50-100 \;{\rm km}$ before decaying. Compared to the system scale they decay more or less where they are produced and their daughter particles will produce their full thick target yield of bremsstrahlung photons. Radioactive positron-emitter nuclei can have lifetimes of several minutes and longer but, as noted earlier, are not included in our simulations here.

In the simulations carried out in this work we consider primary accelerated ions both with ``photospheric'' composition and with ``impulsive-flare'' composition as defined in \cite{Mandzhavidze1993},  \cite{Ramaty1996} and \cite{Murphy1997}. For the photospheric composition we assume that the abundances of the primary accelerated ions heavier than $\alpha$-particles relative to protons are the same as for the ambient nuclei relative to H, {\it i.e.} the abundances of \cite{Asplund2009}. For the impulsive-flare composition we assume the enhanced relative abundances for the primary accelerated heavy ions given in \cite{Murphy2007}, as indicated in Table \ref{tab:compAtSolarModIII}, and adopt values of $0.1$ and $0.5$ for the $\alpha$/proton ratio. We do not include $^3{\rm He}$ in the compositions of the ambient nuclei and the primary accelerated heavy ions, although it may be important in various ways for a full interpretation of the $\gamma$-ray spectrum \cite[]{Murphy2016}. In any case FLUKA's treatment of reactions involving $^3$He is incomplete at present. One should also note that we assume the same primary accelerated heavy ion abundances for both values of the $\alpha$/proton ratio, although according to \cite{Murphy2007} in the case where the $\alpha$/proton ratio is $0.1$ the abundances should be reduced by a factor of $5$. Nevertheless, one should note that our fits to observed $\gamma$-ray spectra from solar flares have been carried out by using two separate templates for the components due to the direct and the inverse reactions, such that the primary accelerated heavy ion abundances are actually multiplied by a common factor which depends on the yields of the direct and the inverse components and their normalizations determined by the fitting procedure (see Equation~\ref{factor}).

We run separate simulations for the reactions involving each primary accelerated ion-species $i$ and all ambient nuclei. The primary accelerated ions are assumed to have power-law energy distributions given by:
\begin{equation}
\frac{{\rm d}n_i(E)}{{\rm d}E}=w_i N E^{-\delta} H(E_{\rm max}-E)H(E-E_{\rm min}) \; ,
\label{power-law:ions}
\end{equation}
\noindent
where $w_i$ is the relative abundance for the ion-species $i$, $E$ is the primary accelerated ion kinetic energy {\it per} nucleon in the range from $E_{\rm min}=1 \; {\rm MeV/nucleon}$ to $E_{\rm max}=1 \; {\rm GeV/nucleon}$, $\delta$ is the power-law spectral index, $H$ is the Heaviside step function, and $N$ is a normalization constant defined such that
\begin{equation}
N\int_{E_{\rm min}}^{E_{\rm max}}E^{-\delta} {\rm d}E =1\; .
\end{equation}

FLUKA allows the user to customise various aspects of the simulation by changing the code in certain subroutines. We modified the subroutine source.f  so that it either draws primary ion energies from the power-law distribution Equation~\ref{power-law:ions}, or equivalently draws them randomly from a uniform distribution but assigns weights given by the power-law distribution \cite[]{MacKinnon2016}. The latter strategy accounts for the contribution of the higher-energy particles to the photon spectrum without suffering large statistical fluctuations resulting from their comparative rarity. We generally found that $10^7 - 10^8$ primaries were necessary for a statistically reliable spectrum. Our modified version of source.f provides a number of options for primary ion angular distribution \cite[]{MacKinnon2016} but in this work we always assume that ion directions are drawn from a uniform distribution in the downward hemisphere (``downward isotropic'').

The algorithm implemented by FLUKA follows the evolution of primary and secondary particles individually as they propagate through the target, tracking their interactions with the ambient medium until they leave the target region, come to rest, or reach a low-energy threshold for transport. Several kinds of estimation tools are available, referred to as ``detectors'', which allow one to calculate different quantities of interest. In the simulations carried out with the model described above we use a specific detector, named USRBDX, to determine the energy spectrum of photons escaping from the chromospheric/photospheric region to the coronal region. This kind of detector calculates the double differential distribution of the flux of photons crossing the $xy$-plane at $z=0$ {\it per} energy and solid angle intervals, ${\rm d}^2 \phi/{\rm d}E {\rm d}\Omega$, in units of photons ${\rm GeV}^{-1}\;{\rm cm}^{-2}\;{\rm sr}^{-1}$ {\it per} primary accelerated ion. Integrating the flux of photons in the upward direction over the solid angle and the whole of the surface area defined by the $xy$-plane at $z=0$, we obtain the corresponding energy spectrum of escaping photons, ${\rm d}\phi/{\rm d}E$, in units of photons ${\rm GeV}^{-1}$ {\it per} primary accelerated ion -- {\it i.e.} the photon spectrum emitted into the backward hemisphere (coronal region). The user determines the energy resolution of the detector by specifying the number of bins and the maximum and minimum energies of photons to be recorded. We chose these by experiment to obtain a reasonable compromise between spectral resolution and statistical reliability. We also set additional detectors with narrower, $3 \; {\rm keV}$ energy bins to give better resolution around the strong $511 \; {\rm keV}$ and $2.223\; {\rm MeV}$ lines, combining these with the full spectrum in the results shown below. In Section \ref{S-Lines} we compare spectra of nuclear de-excitation $\gamma$-ray lines obtained in simulations with FLUKA and in calculations carried out with a copy of the RMK code kindly supplied by R. J. Murphy. Because the RMK calculated spectra are obtained by integrating the flux of photons in all directions, spectra calculated with FLUKA are expected to be nearly a factor of $2$ less intense than those calculated with the RMK code. Thus, for a proper comparison of the spectra calculated with the two codes, in the figures shown we will decrease the RMK spectra by a factor two.

As pointed out before, the contributions to the solar flare $\gamma$-ray spectrum from the emission of nuclear de-excitation, neutron capture and positron annihilation lines are produced mostly by primary accelerated ions with energies in the range from $\approx 1$ to $100 \;{\rm MeV/nucleon}$. The continuum emission contributions from pion decay processes, on the other hand, are produced by primary ions with energies above $\approx 200-300 \; {\rm MeV/nucleon}$. Thus, in order to improve the statistics for the photon yields generated by all emission processes we run separate simulations for two energy ranges, namely $E_{\rm min}^{\rm low}=1 \;{\rm MeV/nucleon}$ to $E_{\rm max}^{\rm low}=200 \;{\rm MeV/nucleon}$ and $E_{\rm min}^{\rm high}=200 \; {\rm MeV/nucleon}$ to $E_{\rm max}^{\rm high}=1 \; {\rm GeV/nucleon}$. The energy spectrum of photons, ${\rm d}\phi_i(E)/{\rm d}E$, for a given ion-species $i$ is obtained by summing up the contributions from the two energy ranges with the appropriate weighting factors,
\begin{equation}
\frac{{\rm d}\phi_i(E)}{{\rm d}E}=N_{\rm low}\frac{{\rm d}\phi_{i}^{\rm low}(E)}{{\rm d}E}+N_{\rm high}\frac{{\rm d}\phi_{i}^{\rm high}(E)}{{\rm d}E}
\end{equation}
\noindent
where the weights $N_{\rm low}$ and $N_{\rm high}$ are given by:
\begin{eqnarray}
N_{\rm low}=N\int_{E_{\rm min}^{\rm low}}^{E_{\rm max}^{\rm low}}E^{-\delta}{\rm d}E \; ,\\ \nonumber\\
N_{\rm high}=N\int_{E_{\rm min}^{\rm high}}^{E_{\rm max}^{\rm high}}E^{-\delta}{\rm d}E \; .
\end{eqnarray}
\noindent
Thus, by summing up the contributions from each ion-species $i$ weighted by their corresponding relative abundances, $w_i$, we obtain the total energy spectrum of photons in units of photons ${\rm GeV}^{-1}$ {\it per} primary accelerated proton in the energy range from $E_{\rm min}$ to $E_{\rm max}$,
\begin{equation}
\frac{{\rm d}\phi(E)}{{\rm d}E}=\sum_i w_i \frac{{\rm d}\phi_i(E)}{{\rm d}E} \; .
\end{equation}

Similarly, we can obtain the energy spectra of photons produced by the direct reactions (involving interactions of primary accelerated protons and $\alpha$-particles with all ambient nuclei) and by the inverse reactions (involving interactions of primary accelerated ions heavier than $\alpha$-particles with ambient nuclei of H and $^4$He), given by
\begin{equation}
\frac{{\rm d}\phi_{\rm dir}(E)}{{\rm d}E}= w_p \frac{{\rm d}\phi_p(E)}{{\rm d}E}+w_{\alpha} \frac{{\rm d}\phi_{\alpha}(E)}{{\rm d}E} \; ,
\label{dir-spec}
\end{equation}
\begin{equation}
\frac{{\rm d}\phi_{\rm inv}(E)}{{\rm d}E}=\sum_{i\neq p,\alpha} w_i \frac{{\rm d}\phi_i(E)}{{\rm d}E} \; .
\label{inv-spec}
\end{equation}

Since the RMK spectra are normalized to one primary accelerated proton in the energy range from $30 \; {\rm MeV}$ to $1 \; {\rm GeV}$, we renormalize the FLUKA spectra of nuclear de-excitation $\gamma$-ray lines in order to allow for a proper comparison, {\it i.e.},
\begin{eqnarray}
\frac{{\rm d}\phi_{\rm dir}(E)}{{\rm d}E}&\rightarrow&\frac{N_{30}}{N}\frac{{\rm d}\phi_{\rm dir}(E)}{{\rm d}E} \\ \nonumber\\
\frac{{\rm d}\phi_{\rm inv}(E)}{{\rm d}E}&\rightarrow&\frac{N_{30}}{N}\frac{{\rm d}\phi_{\rm inv}(E)}{{\rm d}E}
\end{eqnarray}
\noindent
where $N_{30}$ is the normalization constant for the power-law energy distributions considered in the calculations with the RMK code, given by:
\begin{equation}
N_{30}=\left[\int_{30~{\rm MeV}}^{1~{\rm GeV}}E^{-\delta}{\rm d}E\right]^{-1}\; .
\end{equation}
\noindent
The same renormalization was carried out for the full $\gamma$-ray spectra obtained in simulations with FLUKA shown in Section \ref{S-Full}.

%%%%%%%%%%%%%%%%%%%%%%%%%%%%%%%%%%%%%%%%%%%%%%%%%%%%%%%%%%

\section{Results of Simulations}
\label{S-simulations}

In this section we present the results obtained in simulations carried out with FLUKA using the geometrical model for the ambient solar atmosphere described above. In Section \ref{S-Lines} we compare the nuclear de-excitation $\gamma$-ray line spectra obtained in simulations with FLUKA and in calculations with the RMK code. In Section \ref{S-Full} we show the results obtained in simulations with FLUKA for the full $\gamma$-ray spectra, {\it i.e.} with other relevant photon emission processes apart from the positron annihilation line produced via radioactive positron-emitter daughter nuclei taken into account. In all cases, both for RMK and FLUKA spectra, we consider downward isotropic beams of primary accelerated ions with power-law energy distribution of spectral index $\delta=4$ in the range from $1 \; {\rm MeV/nucleon}$ to $1 \; {\rm GeV/nucleon}$.

\subsection{Nuclear De-excitation $\gamma$-Ray Line Spectrum}
     \label{S-Lines}

The RMK code \cite[]{Ramaty1979, Kozlovsky2002, Murphy2009} is one of the main tools used for the analysis of solar flare $\gamma$-ray data. The code is based on a Monte Carlo algorithm which allows one to calculate the ``explicit'' narrow and broad nuclear de-excitation $\gamma$-ray lines and the so called ``unresolved'' component. By ``explicit'' line is meant one of the more than $200$ lines that are explicitly treated by the code. These are the strongest lines resulting from transitions of the excited nuclei produced by interactions of the most abundant nuclei in the solar atmosphere. The unresolved component is composed of the thousands of remaining lines produced by the interactions which are relatively weak and closely spaced. They are treated by the code as a ``quasi-continuum''. The nuclear de-excitation $\gamma$-ray line spectra generated with the RMK code are evaluated by integrating over solid angle the flux of photons in all directions and take into account the dependence of the Doppler-shifts and the relativistic beaming of moving emitter nuclei on the heliocentric angle $\theta_{\rm obs}$. The updated version of the code \cite[]{Murphy2009} provides the option to include or not the effects of Compton scattering. The RMK spectra shown in this work are evaluated with Compton scattering turned off.

The production of nuclear de-excitation $\gamma$-ray lines in FLUKA occurs at the final stage of a sequence of nuclear processes implemented through a model called PEANUT (Pre-Equilibrium Approach to Nuclear Thermalization), developed by \cite{Ferrari1998} (see the Appendix for details). In order to concentrate on the nuclear de-excitation $\gamma$-ray line spectra we use the functionalities of FLUKA which allow one to suppress the processes of bremsstrahlung, production and annihilation of electron-positron pairs, neutron capture, Compton scattering and pion decay. The FLUKA calculations include production of de-excitation-lines from excited nuclei produced by inelastic excitation reactions and from spallation reactions. We do not include fusion reactions in the FLUKA calculations. Fusion reactions are included in the RMK calculations, but apart from the $\alpha$-$\alpha$ fusion lines near $0.45 \; {\rm MeV}$ and $3.6 \; {\rm MeV}$, fusion reactions do not play a major role in the spectra of nuclear de-excitation $\gamma$-ray lines. It is possible to enable fusion reactions in FLUKA and model the $\alpha$-$\alpha$ feature if desired.

The nuclear de-excitation $\gamma$-ray line spectra generated in our FLUKA simulations (as well as the full $\gamma$-ray spectra) are evaluated by integrating over solid angle only the flux of photons in the upward direction ({\it i.e.}, the flux of photons escaping from the chromospheric/photospheric region to the coronal region). In this way, for downward isotropic beams of primary accelerated ions we obtain photon spectra which correspond to an average of spectra for flares at  heliocentric angles between $\theta_{\rm obs} = 0^{\circ}$ (disk-centered) and $\theta_{\rm obs}=90^{\circ}$ (limb) and so exhibit Doppler-shifts similar to that of a spectrum for a flare at $\theta_{\rm obs} = 60^{\circ}$, considering that the magnitude of the Doppler-shifts roughly changes linearly with ${\rm cos} (\theta_{\rm obs})$ from $0$ for a limb flare to a minimum negative value for a disk-centered flare \cite[]{Harris2007}. Thus, in order to perform a more consistent comparison between the nuclear de-excitation $\gamma$-ray line spectra obtained in simulations with FLUKA and in calculations with the RMK code, we generate the RMK spectra for $\theta_{\rm obs} = 60^{\circ}$. Photon spectra which approximately correspond to spectra for specific heliocentric angles $\theta_{\rm obs}$ could be obtained in our FLUKA simulations by setting up a high-resolution solid angle binning and integrating over very small solid angle intervals, {\it cf.} the angle-dependent spectra at higher energies obtained using GEANT4 by \cite{Tang2010}. We would have to substantially increase run times to obtain such angle-dependent spectra with good statistics and small enough energy bins to see the effects on line shapes, however, and we leave such an effort to future work.

\begin{figure}[ht]
    \centering
    \includegraphics[width=0.95\linewidth]{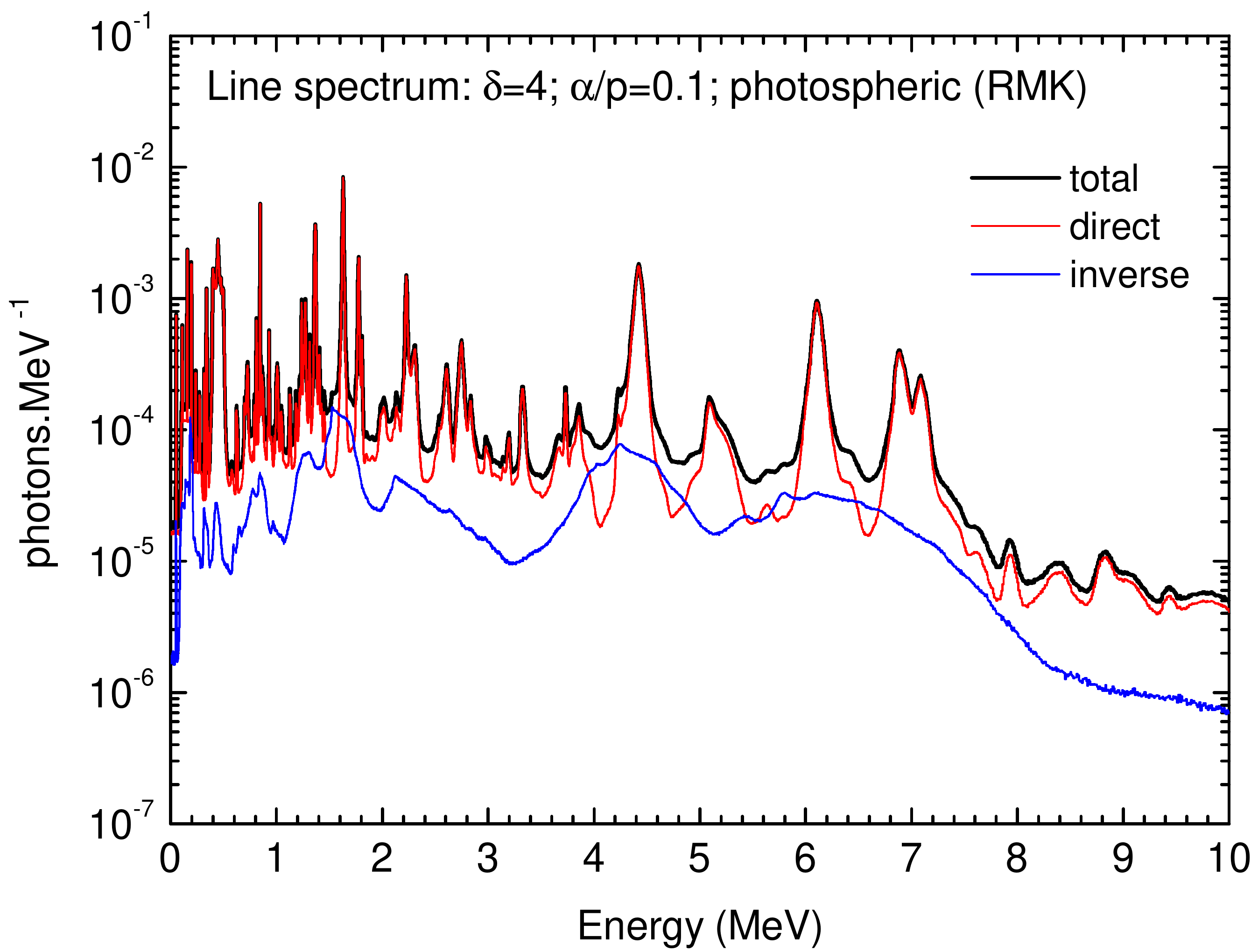}\\
    \includegraphics[width=0.95\linewidth]{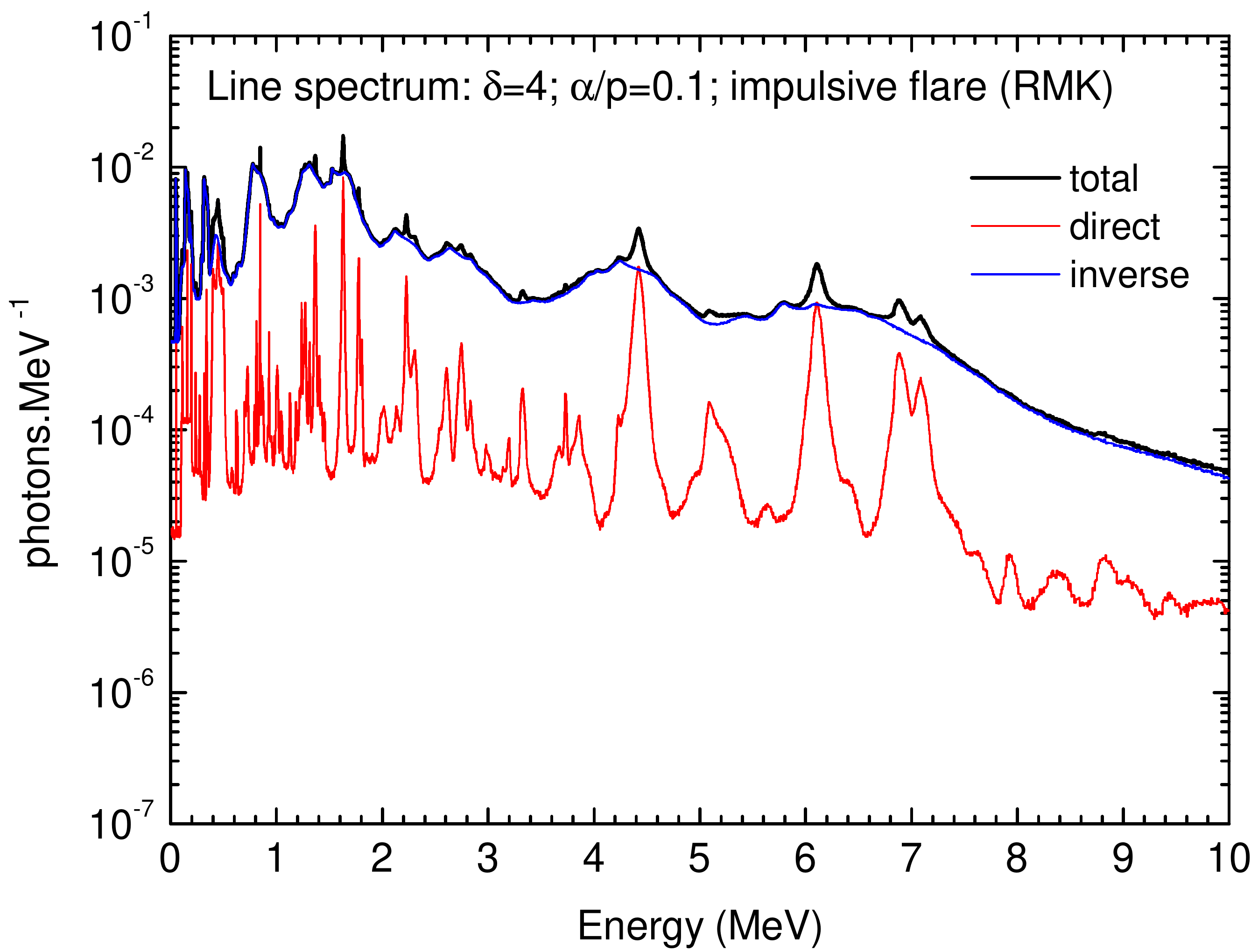}
  \caption{Nuclear de-excitation $\gamma$-ray line spectra obtained in calculations with the RMK code for primary accelerated ions with photospheric and impulsive-flare compositions.}
   \label{F-RMK-Lines-D4}
\end{figure}

\begin{figure}[ht]
    \centering
    \includegraphics[width=0.95\linewidth]{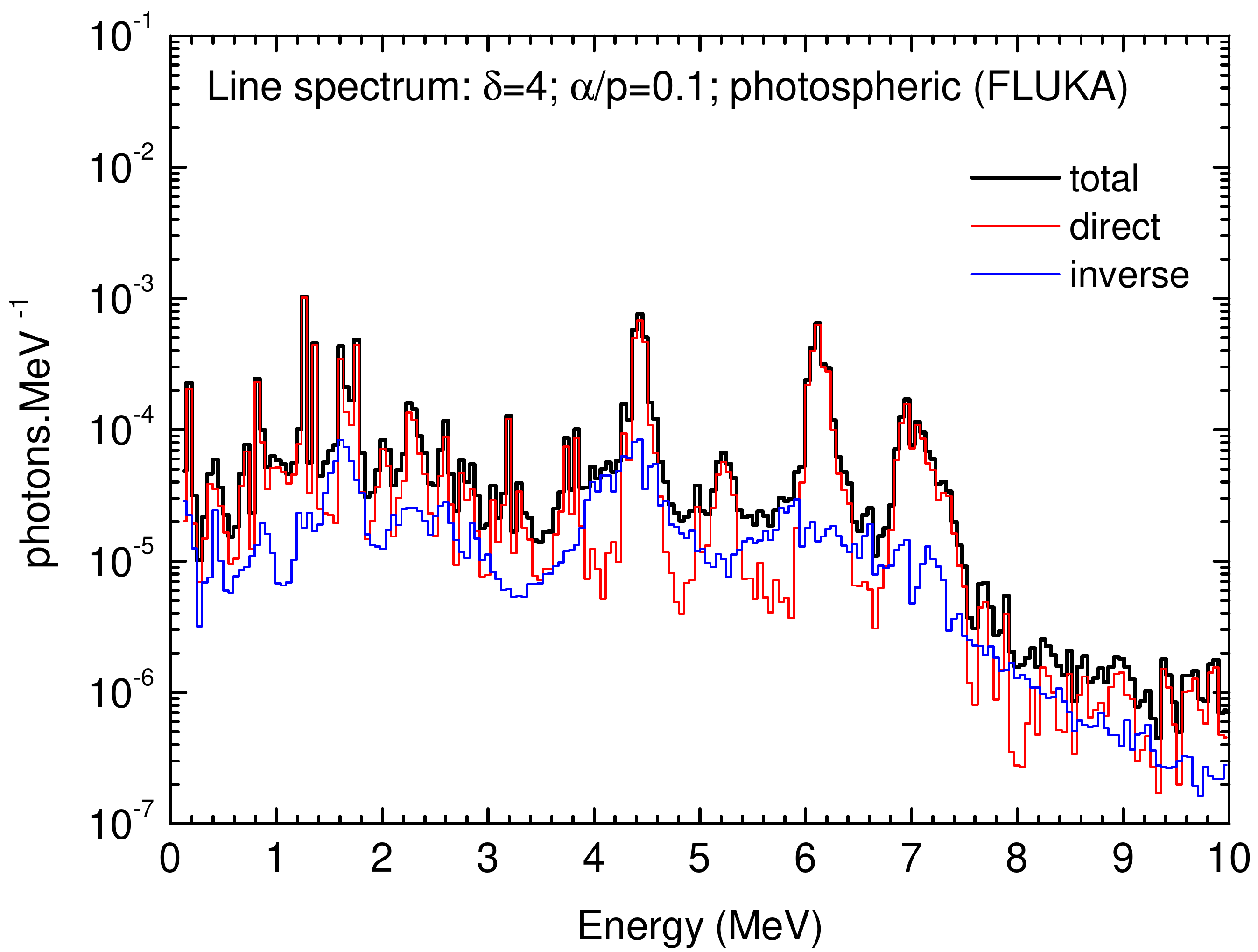}\\
    \includegraphics[width=0.95\linewidth]{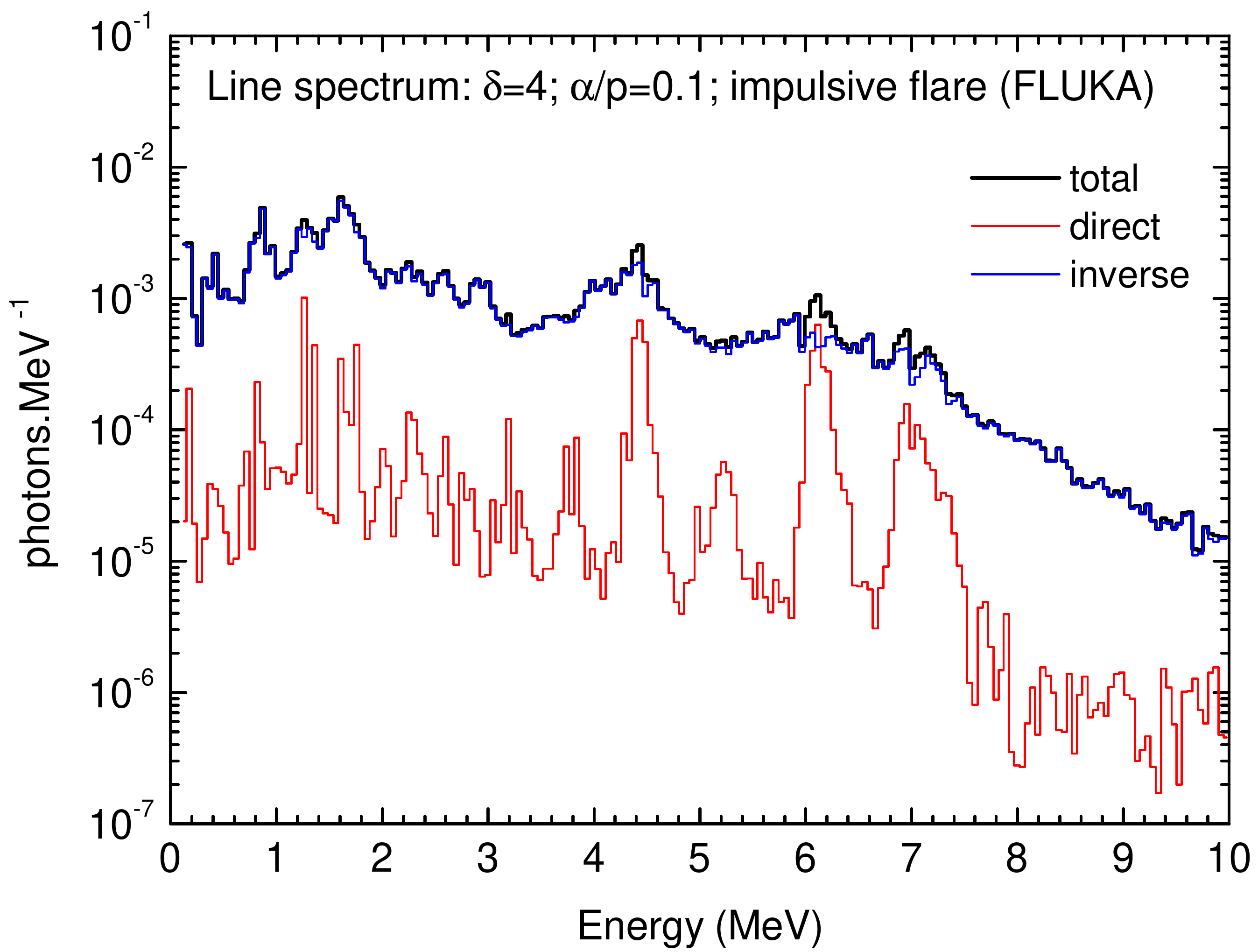}
  \caption{Nuclear de-excitation $\gamma$-ray line spectra obtained in simulations with FLUKA for primary accelerated ions with photospheric and impulsive-flare compositions.}
   \label{F-FLUKA-Lines-D4}
\end{figure}

In Figures \ref{F-RMK-Lines-D4} and \ref{F-FLUKA-Lines-D4} we show nuclear de-excitation $\gamma$-ray line spectra obtained, respectively, in calculations with the RMK code and in simulations with FLUKA for downward isotropic beams of primary accelerated ions with photospheric and impulsive-flare compositions (assuming $\alpha/{\rm proton}=0.1$ and $\delta=4$ in both cases).

\clearpage
\newpage
\pagebreak

\clearpage
\newpage
\pagebreak

\clearpage
\newpage
\pagebreak

\begin{figure}[ht]
    \centering
    \includegraphics[width=0.95\linewidth]{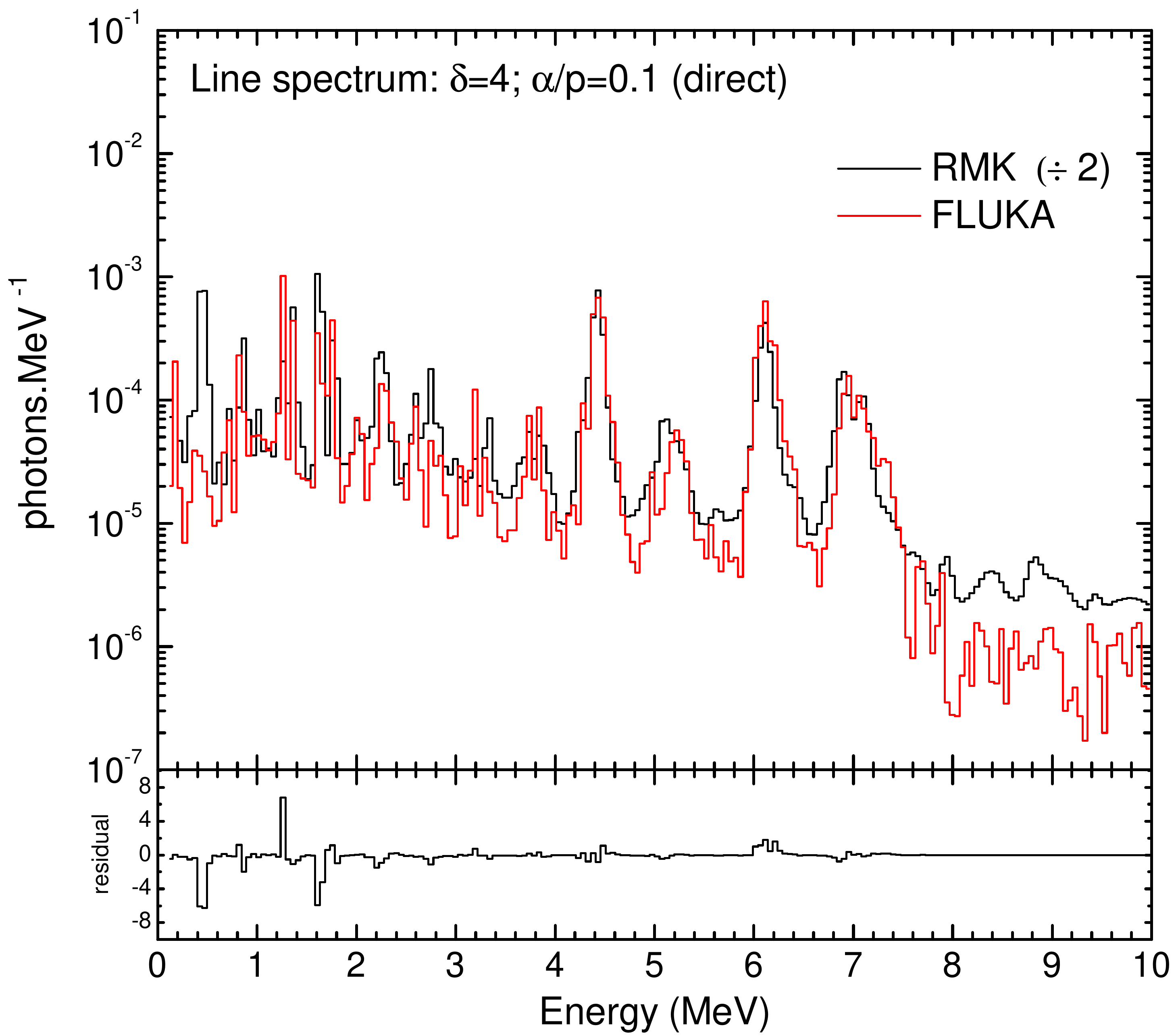}
  \caption{Contributions from direct reactions to the nuclear de-excitation $\gamma$-ray line spectrum obtained in simulations with FLUKA and in calculations with the RMK code.}
   \label{F-RMK-FLUKA-Lines-D4-direct}
\end{figure}

\begin{figure}[ht]
    \centering
    \includegraphics[width=0.95\linewidth]{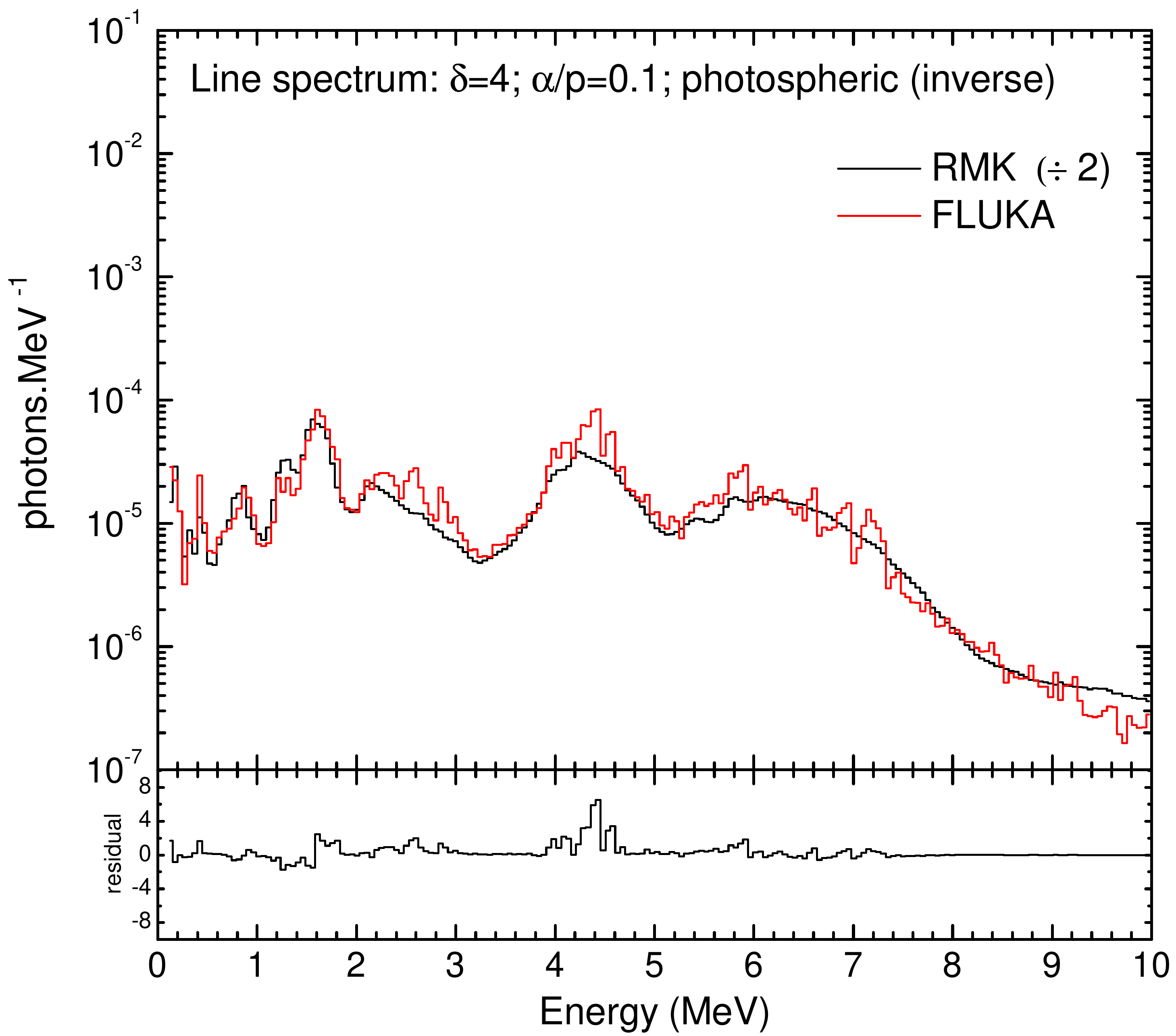}\\
    \includegraphics[width=0.95\linewidth]{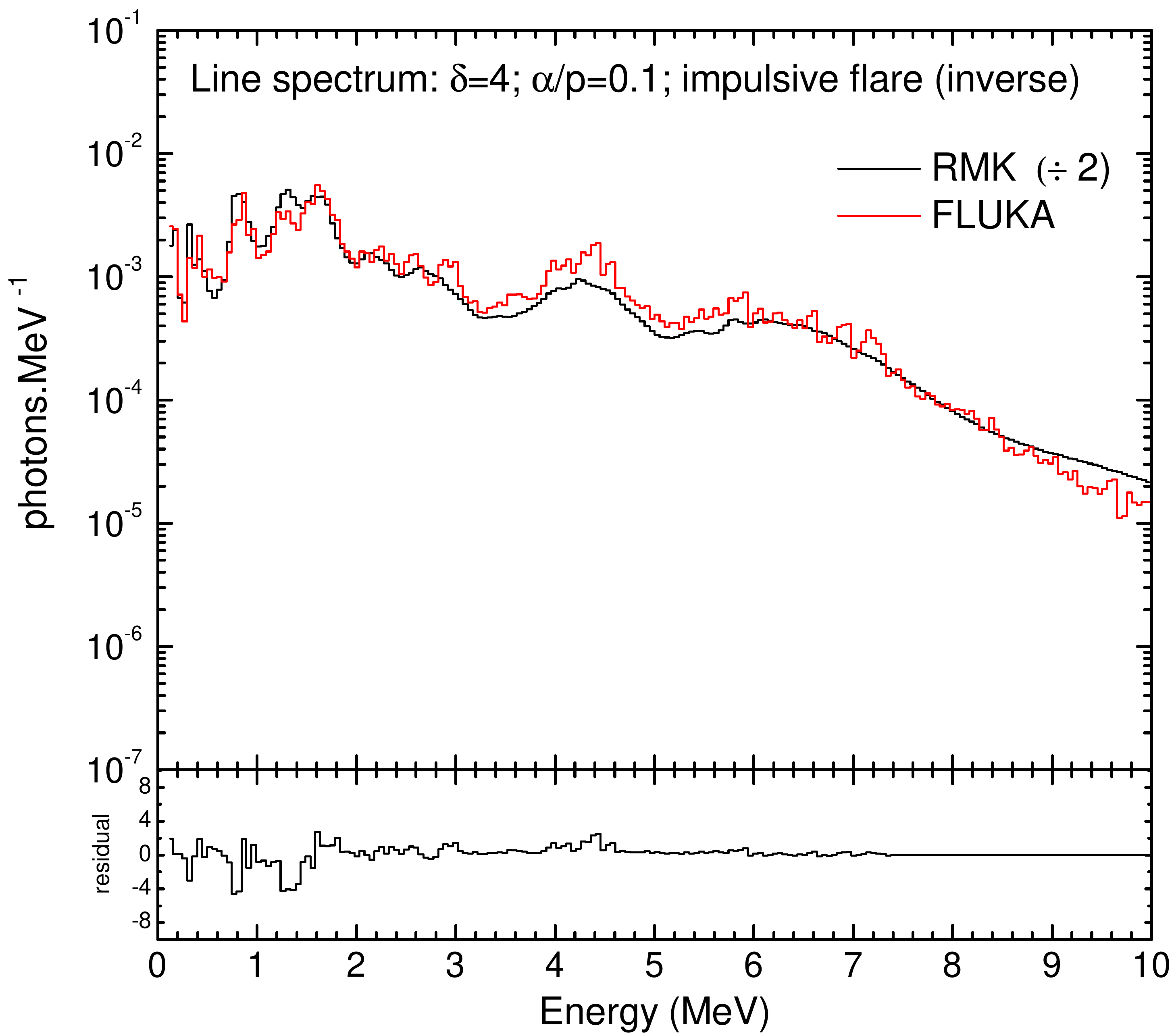}
  \caption{Contributions from inverse reactions to the nuclear de-excitation $\gamma$-ray line spectrum obtained in simulations with FLUKA and in calculations with the RMK code for primary accelerated ions with photospheric and impulsive-flare compositions.}
   \label{F-RMK-FLUKA-Lines-D4-inverse}
\end{figure}

\begin{figure}[ht]
    \centering
    \includegraphics[width=0.95\linewidth]{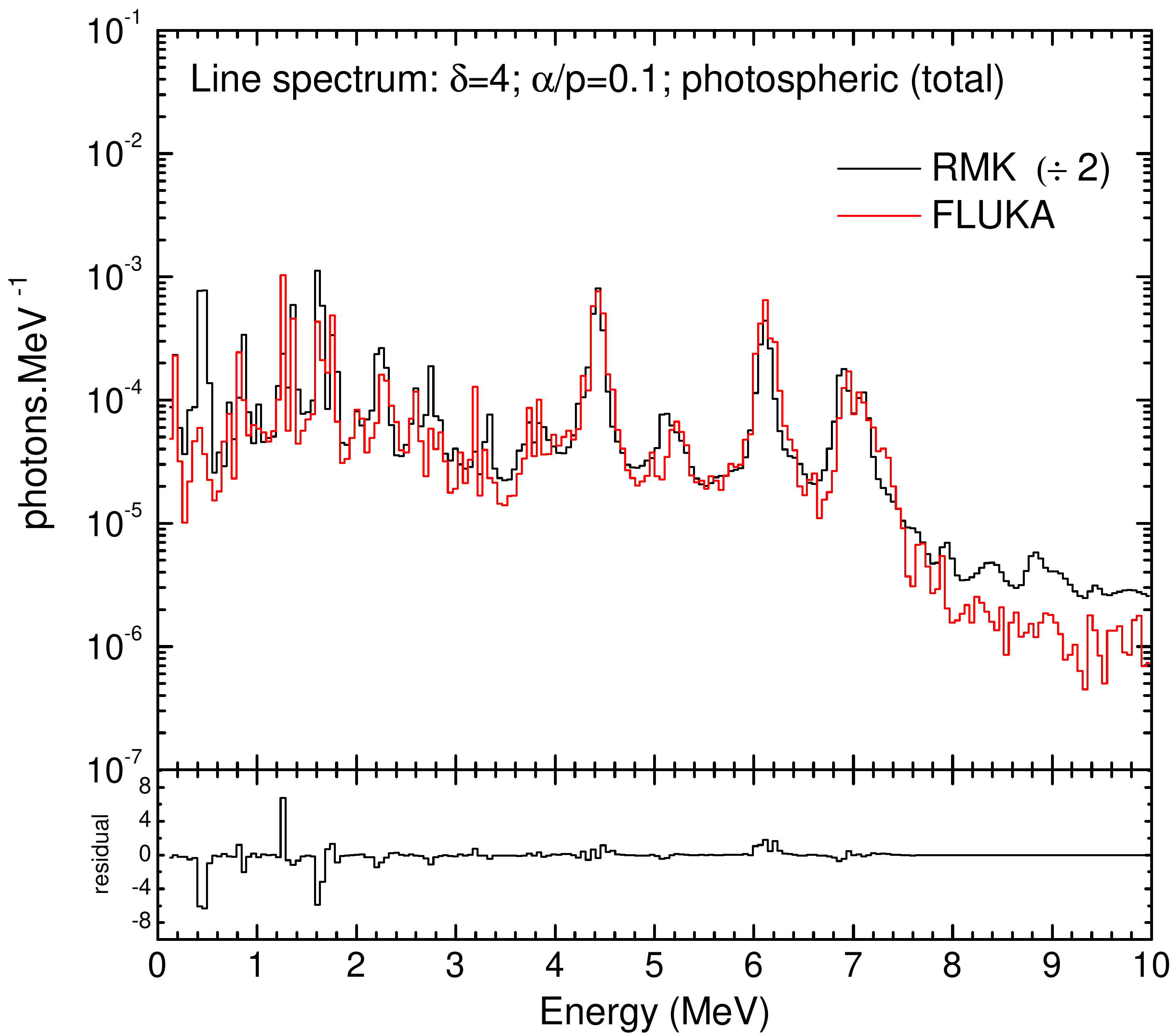}\\
    \includegraphics[width=0.95\linewidth]{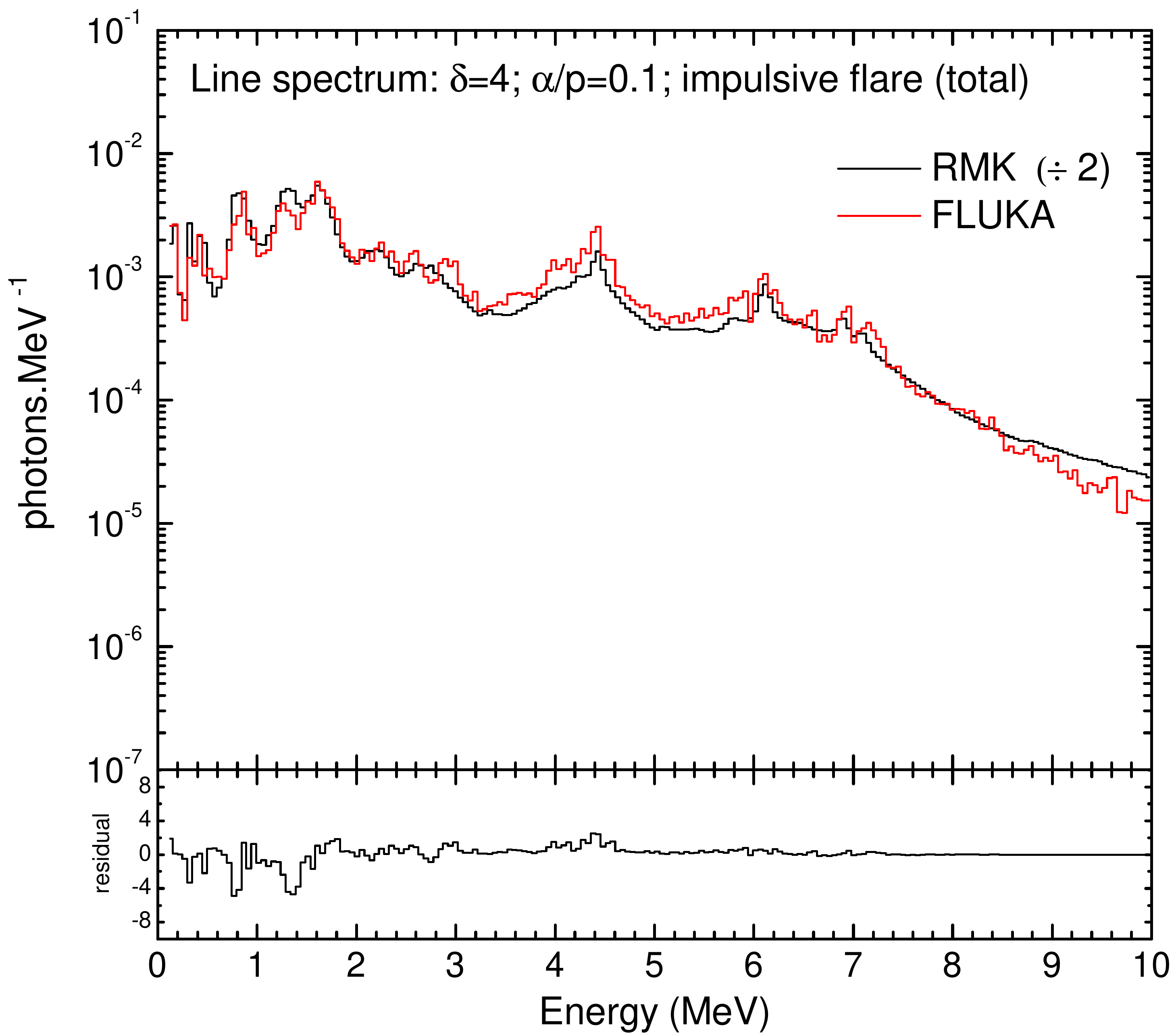}
  \caption{Total nuclear de-excitation $\gamma$-ray line spectra obtained in simulations with FLUKA and in calculations with the RMK code for primary accelerated ions with photospheric and impulsive-flare compositions.}
   \label{F-RMK-FLUKA-Lines-D4-total}
\end{figure}

In Figures \ref{F-RMK-FLUKA-Lines-D4-direct}, \ref{F-RMK-FLUKA-Lines-D4-inverse} and \ref{F-RMK-FLUKA-Lines-D4-total} we compare FLUKA and RMK spectra, showing respectively the contributions from direct and inverse reactions and the total spectrum. For a better comparison, we have rebinned the higher-resolution RMK spectra into the same binning as that of the FLUKA spectra. As one can observe, the spectra obtained in the simulations with FLUKA can satisfactorily reproduce those obtained in calculations with the RMK code. The $0.4 - 0.5 \; {\rm MeV}$ spectral feature from $\alpha - \alpha$ fusion reactions (the superposition of $\gamma$-ray lines produced by de-excitation of $^7$Be and $^7$Li nuclei) does not appear in the FLUKA spectra shown because, as noted above, fusion reactions are not included in the calculations. %There are contributions from spallation reactions to the strong de-excitation lines, which mean that other species than the one the line belongs to can contribute, but these are not fusion reactions.

\clearpage
\newpage
\pagebreak

\clearpage
\newpage
\pagebreak

\clearpage
\newpage
\pagebreak

Figures \ref{F-RMK-FLUKA-Lines-D4-direct} to \ref{F-RMK-FLUKA-Lines-D4-total} show very good agreement between the FLUKA and RMK results, particularly considering that the two codes were developed independently from different starting points. Most of the major line features are present in the FLUKA results, albeit sometimes with poorer statistical quality. In the case of the strongest lines, those produced by de-excitation of $^{12}$C at $4.438 \;{\rm MeV}$ and $^{16}$O at $6.129$, $6.916$ and $7.115 \; {\rm MeV}$, FLUKA also gives similar results for line widths and shapes. The normalised residuals displayed in the bottom panels of the figures show the differences between the FLUKA and RMK photon spectra. In the $i$th energy bin we calculate the difference between the photon fluxes produced by the two codes,
\begin{equation}
r_i \, = \, \frac{{\rm d}\phi(E)}{{\rm d}E}\bigg|_{i, {\rm RMK}} - \frac{{\rm d}\phi(E)}{{\rm d}E}\bigg|_{i,{\rm FLUKA}} \;,
\end{equation}
\noindent
and divide by the standard deviation $\sigma$,
\begin{equation}
\sigma \, = \, \sqrt{\frac{1}{N-1} \sum_i r_i^2} \; ,
\end{equation}
\noindent
where $N$ is the total number of photon energy bins. The residuals focus attention on a few noteworthy discrepancies. First, as already noted, we expect disagreement around $\approx 0.4 - 0.5$ MeV because the $\alpha$-$\alpha$ lines are not included. This omission could be remedied for more detailed comparison with data. Although the shapes of the strongest lines are similar there are some slight differences, partly attributable to FLUKA's treatment of primary ion transport, modelling the detailed development of the initially downward isotropic ion distribution in the atmosphere, but mostly reflecting noise in the FLUKA results. Discrepancies close to the two strongest lines between 1 and 2 MeV, at $1.37$ MeV from $^{24}$Mg and $1.63$ MeV from $^{20}$Ne, are also attributable to statistical fluctuations in the FLUKA spectrum, especially combined with the width of the photon energy bins used in our FLUKA results. Lastly, although the low absolute fluxes mean it is not highlighted by the residuals, the nuclear continuum above $\approx 7$ MeV is systematically substantially lower in the FLUKA results. This reflects the different treatments of the continuum in PEANUT (used in FLUKA) and in TALYS (used in RMK). Further detailed study will be needed to fully resolve this discrepancy. Lacking narrow line features, the contribution from the inverse reactions is less sensitive to the low resolution of the energy binning and all the main features from the inverse reactions shown in the RMK spectra are well reproduced (Figure~\ref{F-RMK-FLUKA-Lines-D4-inverse}).

To make this comparison we ignore attenuation of photons originally moving upward away from the Sun ({\it i.e.}, Compton scattering and pair production above $1.022 \; {\rm MeV}$ are turned off). This is reasonable because the (hydrogen) photon absorption length is greater than $\approx 5 \; {\rm g/cm^2}$ for photon energies $ > 0.5 \;{\rm MeV}$ \cite[]{Berger2010}, whereas protons with energies $< 100 \; {\rm MeV}$ all stop in the atmosphere higher than this, at column-densities $\lesssim 4 \; {\rm g/cm^2}$ \cite[]{Berger1999}. However, Compton down-scattering and escape of $> 1 \; {\rm MeV}$ photons originally moving downward into the Sun does contribute to the spectrum of nuclear de-excitation $\gamma$-ray lines at energies $< 1 \; {\rm MeV}$. In general, this Compton-scattered component has no significant impact on fits to typical $\gamma$-ray spectra from solar flares observed with instruments such as RHESSI which have a strong off-diagonal detector response \cite[]{Murphy2018}. As expected, in the case of beams of primary accelerated ions with impulsive-flare composition the contribution from the inverse reactions significantly increases, since the relative abundance of the heavy ions is enhanced.

\subsection{Full $\gamma$-Ray Spectrum}
     \label{S-Full}

In our FLUKA simulations of the full $\gamma$-ray spectrum we turn on the processes of bremsstrahlung, production and annihilation of electron-positron pairs, neutron capture and pion decay. As previously noted, the contribution from the decay of radioactive positron-emitter daughter nuclei to the $511 \; {\rm keV}$ positron annihilation line is not included but can be accounted for, if desired, by enabling the radioactive decay of nuclei and the transport of decay radiation in the FLUKA simulations. We also turn on Compton scattering, so that the attenuation of the $2.223 \; {\rm MeV}$ neutron capture line and its associated Compton-scattered continuum are taken into account.

\begin{figure}[ht]
    \centering
    \includegraphics[width=0.95\linewidth]{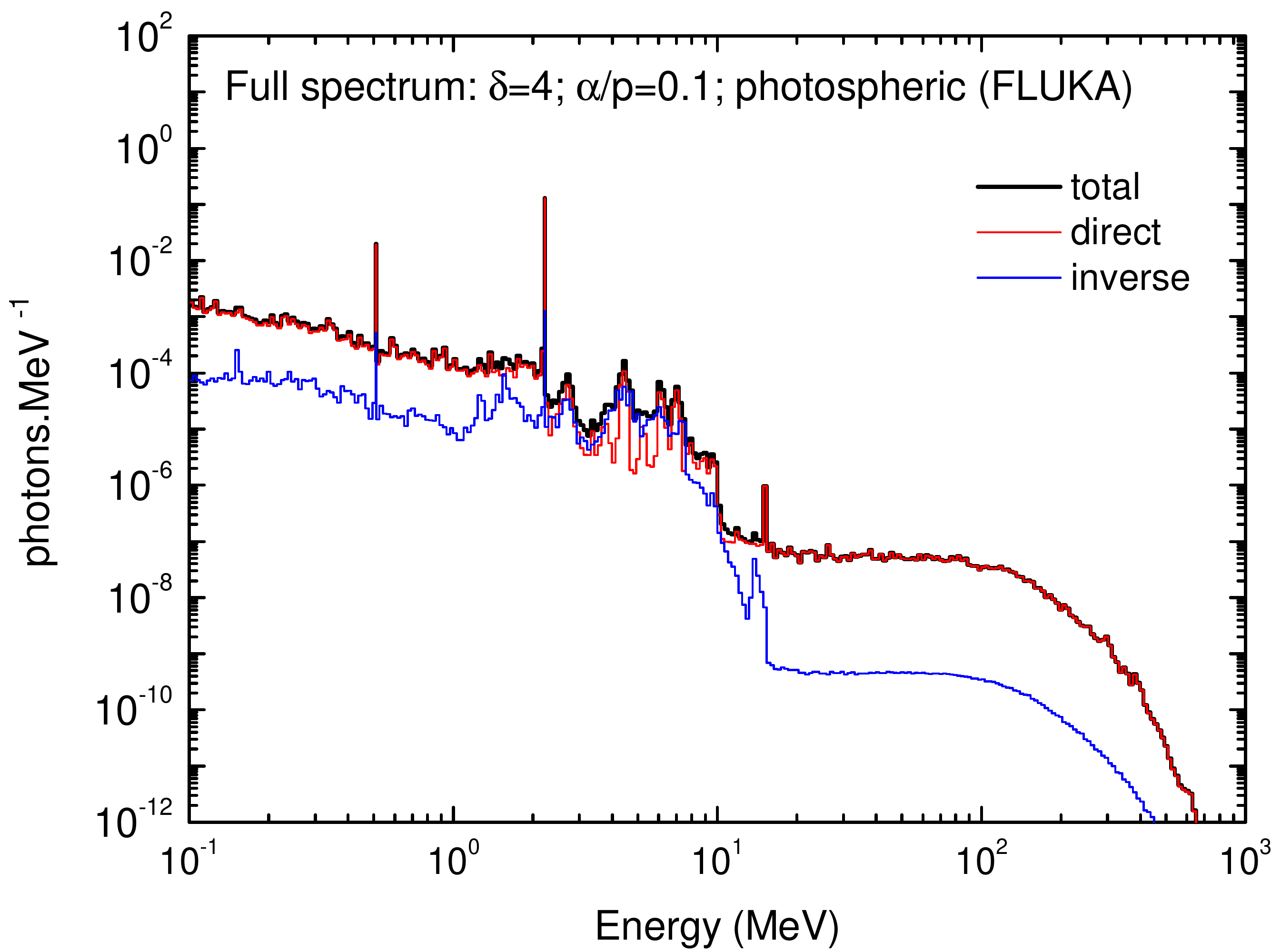}\\
    \includegraphics[width=0.95\linewidth]{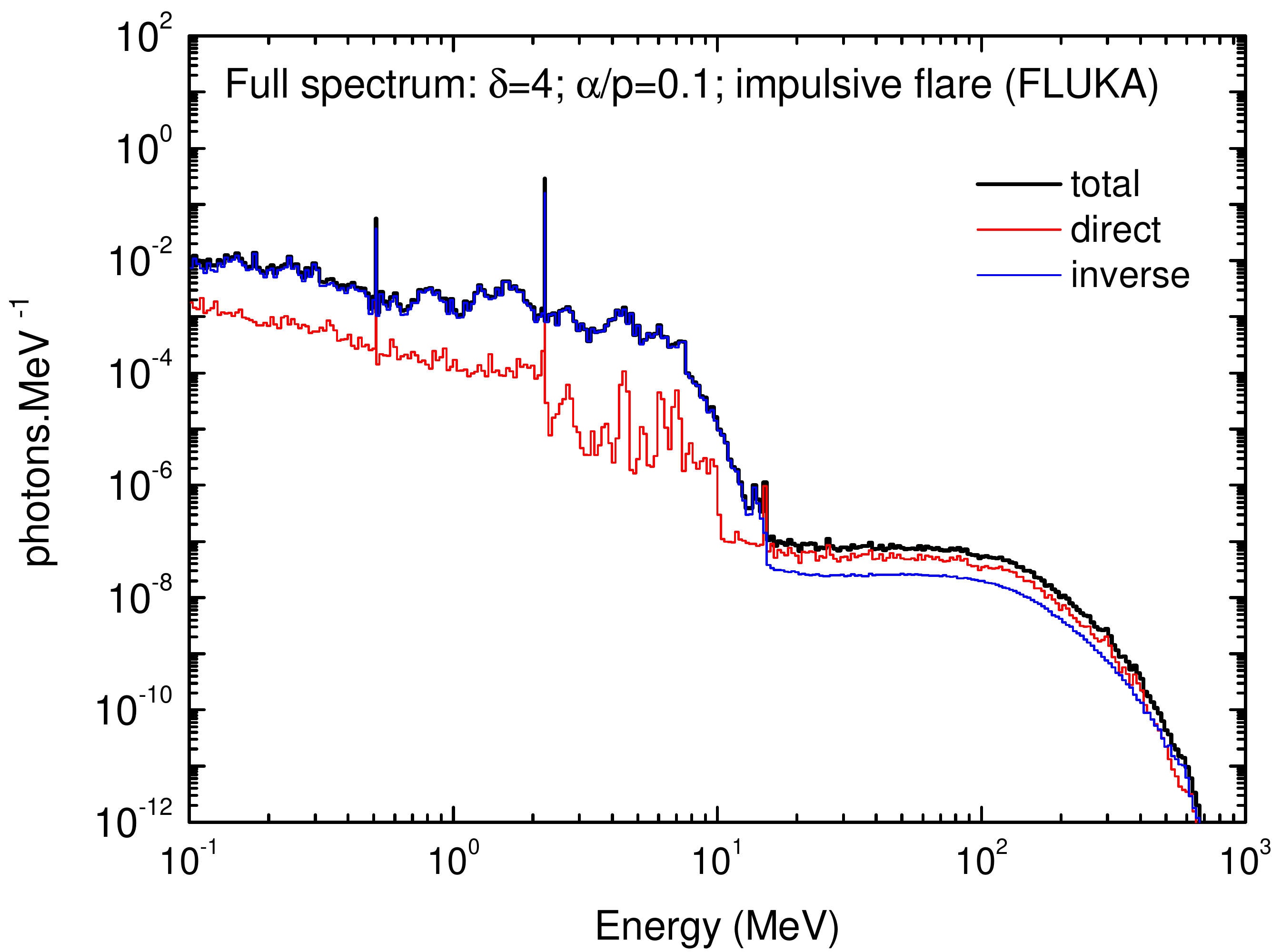}
  \caption{Full $\gamma$-ray spectra obtained in simulations with FLUKA for primary accelerated ions with photospheric and impulsive-flare compositions.}
   \label{F-FLUKA-Full-D4}
\end{figure}

In Figure \ref{F-FLUKA-Full-D4} we show the full $\gamma$-ray spectra obtained for downward isotropic beams of primary accelerated ions with both photospheric and impulsive-flare compositions (assuming $\alpha/{\rm proton}=0.1$ and $\delta=4$ in both cases). As one can observe, the FLUKA spectra exhibit all the typical structures of $\gamma$-ray spectra observed in solar flares: positron annihilation line (though excluding the substantial contribution from radioactive nuclei), neutron capture line, nuclear de-excitation $\gamma$-ray lines and continuum emission components from neutral-pion decay and bremsstrahlung of secondary electrons and positrons from charged-pion decay. One can also see that the component due to the direct reactions clearly exhibits the Compton-scattered continuum from the neutron-capture line at energies below $2.223 \; {\rm MeV}$. This component is intense enough that it conceals the similar Compton component from Compton scattering of the de-excitation lines described by \cite{Murphy2018}. The continuum emission component from the bremsstrahlung of primary accelerated electrons is not included here. As well as the abundances of the heavier species, the intensity of the pion decay continuum from inverse reactions is influenced by the larger cross-sections and lower energy thresholds for pion production in reactions with primary accelerated heavy ions. The relative magnitudes of the inverse and direct reaction fluxes are in general agreement with those found by \cite{Kafexhiu2018}. When spectral features are well-defined enough to allow a direct comparison, {\it e.g.} in the case of the $2.223 \; {\rm MeV}$ neutron capture line, yields from inverse reactions are further diminished by the $Z^2/A$ dependence of the energy loss rate.

\clearpage
\newpage
\pagebreak

\clearpage
\newpage
\pagebreak

We note that the continuum component from Compton scattering of the $2.223 \; {\rm MeV}$ line is much less evident when produced via inverse rather than direct reactions. Although we focus here on total spectra emerging from the atmosphere, FLUKA allows us also to investigate the spatial distribution of particular species or events through the atmosphere. We found that the shorter range of heavier ions, diminished as $Z^2/A$, results in a greater proportion of secondary neutrons capturing higher in the atmosphere, above the $\approx 12 \; {\rm g/cm^2}$ Compton scattering length for $\approx 2 \; {\rm MeV}$ photons. This accounts for the relatively-weaker Compton tail from the inverse reactions.

In the case of beams of primary accelerated ions with photospheric composition the total spectrum for energies below $\approx 2\; {\rm MeV}$ and above $\approx 10 \; {\rm MeV}$ is dominated by the component due to the direct reactions, while in the range from $\approx 2$ to $10\; {\rm MeV}$ the contributions of the components due to direct and inverse reactions are comparable. One should note that for energies below $\approx 2\; {\rm MeV}$ the component due to the direct reactions is enhanced by the Compton-scattered continuum from the $2.223 \; {\rm MeV}$ neutron capture line. In the case of beams of primary accelerated ions with impulsive-flare composition the total spectrum for energies below $\approx 10\; {\rm MeV}$ is dominated by the component due to the inverse reactions, while for energies above $\approx 10\; {\rm MeV}$ the contributions of the components due to direct and inverse reactions are comparable. For the impulsive-flare composition, the contribution from inverse reactions is approximately two orders of magnitude larger than for the photospheric composition, consistent with the substantially enhanced heavy-ion abundances. In both cases, we note that the structures corresponding to the narrow nuclear de-excitation $\gamma$-ray lines in the energy range from $\approx 2$ to $10\; {\rm MeV}$ are very well-defined in the spectrum of the component due to the direct reactions. For energies below $2\; {\rm MeV}$, on the other hand, the spectrum of the component due to the direct reactions is dominated by the continuum emission from the Compton scattering of the $2.223 \; {\rm MeV}$ neutron capture line, such that the structures corresponding to the narrow nuclear de-excitation $\gamma$-ray lines become nearly invisible.

Aiming to compare our results with previous calculations, we extracted the yields of the neutron-capture line at $2.223 \; {\rm MeV}$ and the $^{12}$C de-excitation line at $4.438 \;{\rm MeV}$ from the FLUKA simulation for primary accelerated ions with impulsive flare composition. The yield of the $2.223 \; {\rm MeV}$ line was obtained from the total, full $\gamma$-ray spectrum. For the yield of the $4.438 \;{\rm MeV}$ line, however, we used the direct component of the de-excitation $\gamma$-ray line spectrum, which has better statistics and energy binning resolution. The resulting value obtained for the fluence ratio of the $2.223 \; {\rm MeV}$ line to the $4.438 \;{\rm MeV}$ $^{12}$C line is $8.7$, in remarkably good agreement with the value of $8.4$ calculated by \cite{Murphy2007} for similar model parameters.

\clearpage
\newpage
\pagebreak

\begin{figure}[t]
    \centering
    \includegraphics[width=0.95\linewidth]{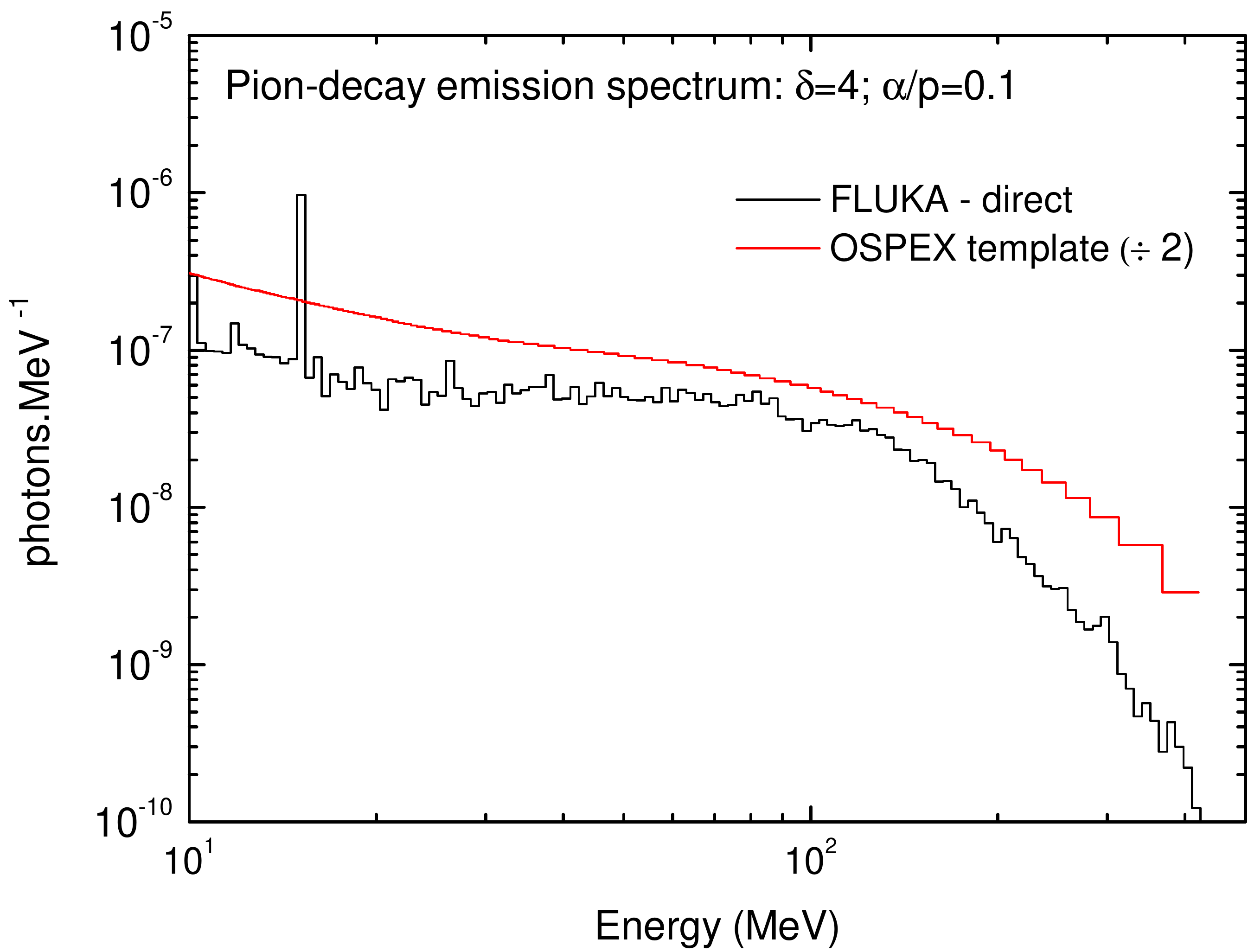}
  \caption{Pion decay emission obtained from the FLUKA simulation of the full $\gamma$-ray spectrum (direct component) and from the OSPEX pion template built from calculations carried out by \cite{Murphy1987}.}
   \label{F-RMK-FLUKA-Pion-D4}
\end{figure}

Lastly, in Figure~\ref{F-RMK-FLUKA-Pion-D4} we show the pion decay continuum above $10 \; {\rm MeV}$ resulting from direct reactions (assuming $\alpha/{\rm proton}=0.1$ and $\delta=4$). For comparison we also show the corresponding spectral template supplied in OSPEX, based on \cite{Murphy1987}. We consider the two spectra to be in reasonable agreement given: 1. the OSPEX template is calculated for primary particles extending to $10 \; {\rm GeV/nucleon}$ rather than the $1 \;{\rm GeV/nucleon}$ used here; 2. since the FLUKA results are integrated only over the backward hemisphere we reduced the OSPEX spectrum by a factor of 2, as discussed in Section~\ref{s-model}, but some modest anisotropy will still reduce the backward hemisphere spectrum compared to an average over $4\pi$ steradians; 3. the FLUKA calculation includes Compton scattering of photons before they leave the solar atmosphere; 4. the FLUKA spectrum also shows the 15.1 MeV $^{12}$C line which is not included in the OSPEX pion decay template. A more detailed discussion of high-energy continuum, directionality, {\it etc}. will be provided in \cite{MacKinnon2016}.

\section{Fits to Observed $\gamma$-Ray Spectra}
\label{S-fitting}

In this section we present the results for the fit to the $\gamma$-ray emission spectrum of the 2010 June 12 solar flare obtained using templates built from simulations carried out with FLUKA for the full $\gamma$-ray spectrum, which are incorporated into OSPEX. The fit is implemented by combining the FLUKA templates with standard functions available in OSPEX to account for the continuum emission from the bremsstrahlung of primary accelerated electrons. We consider downward isotropic beams of primary accelerated ions with power-law energy distribution of spectral index $\delta=4$ in the range from $1 \; {\rm MeV/nucleon}$ to $1 \;{\rm GeV/nucleon}$ and impulsive-flare composition (assuming $\alpha/{\rm proton}=0.1$).

Fits to $\gamma$-ray spectra from solar flares have been performed either by using two separate templates for the components of the nuclear de-excitation $\gamma$-ray line spectrum due to the direct reactions (narrow lines) and the inverse reactions (broad lines) or a single template for both components. For the fit to the spectrum of the 2010 June 10 solar flare, \cite{Ackermann2012} used a single template built with the RMK code. Here we choose to perform the fit by using two separate templates. It is important to note that in the case of fits performed with a single template the ratio between the contributions of the direct and inverse components to the total best-fit model spectrum is fixed, while in the case of fits performed with two separate templates the ratio between the contributions is determined by the fitting procedure. One should also note that although the FLUKA spectra lack the contribution to the $511 \; {\rm keV}$ line of radioactive positron emitters, we do not add a gaussian line component to the spectrum at this energy. The line is barely evident above the electron bremsstrahlung continuum in the GBM data we will analyze, and inclusion of a further line component would not significantly improve the fit to data.

As pointed out in Section \ref{s-model}, the photon spectra used to build the FLUKA templates are renormalized to one primary accelerated proton in the energy range from $30 \; {\rm MeV}$ to $1 \; {\rm GeV}$. Moreover, in order to perform the fit with OSPEX\footnote{see \url{https://hesperia.gsfc.nasa.gov/ssw/packages/xray/dbase/nuclear_template_normalization_readme.txt}.} the FLUKA templates are further renormalized to one photon in the energy range considered, {\it i.e.},
\begin{eqnarray}
\frac{{\rm d}\phi_{\rm dir}(E)}{{\rm d}E}\rightarrow \frac{{\rm d}{\tilde \phi}_{\rm dir}(E)}{{\rm d}E} &=& \frac{1}{I_{\rm dir}}\left[w_p \frac{{\rm d}\phi_p(E)}{{\rm d}E}+w_{\alpha} \frac{{\rm d}\phi_{\alpha}(E)}{{\rm d}E} \right]\; ,\\ \nonumber\\
\frac{{\rm d}\phi_{\rm inv}(E)}{{\rm d}E}\rightarrow \frac{{\rm d}{\tilde \phi}_{\rm inv}(E)}{{\rm d}E} &=&\frac{1}{I_{\rm inv}}\left[\sum_{i\neq p,\alpha} w_i \frac{{\rm d}\phi_i(E)}{{\rm d}E}\right] \; .
\end{eqnarray}
\noindent
where $I_{\rm dir}$ and $I_{\rm inv}$ are respectively the yields of the direct and the inverse components, given by:
\begin{eqnarray}
I_{\rm dir} = \int_{E_{\rm min}^{\rm phot}}^{E_{\rm max}^{\rm phot}}\frac{{\rm d}\phi_{\rm dir}(E)}{{\rm d}E} {\rm d}E\;,\\ \nonumber\\
I_{\rm inv} = \int_{E_{\rm min}^{\rm phot}}^{E_{\rm max}^{\rm phot}}\frac{{\rm d}\phi_{\rm inv}(E)}{{\rm d}E} {\rm d}E \;.
\end{eqnarray}
\noindent
The FLUKA templates for the full $\gamma$-ray spectra cover the energy range from $E_{\rm min}^{\rm phot}=100 \; {\rm keV}$ to $E_{\rm max}^{\rm phot}=1 \; {\rm GeV}$.

The sum of the contributions of the direct and inverse components to the total best-fit model spectrum is given by:
\begin{eqnarray}
\frac{{\rm d}{\tilde \phi}_{\rm tot}(E)}{{\rm d}E}= a[0]_{\rm dir}\frac{{\rm d}{\tilde \phi}_{\rm dir}(E)}{{\rm d}E} + a[0]_{\rm inv}\frac{{\rm d}{\tilde \phi}_{\rm inv}(E)}{{\rm d}E} \; ,
\label{sum-inv-dir}
\end{eqnarray}
\noindent
where $a[0]_{\rm dir}$ and $a[0]_{\rm inv}$ are respectively the normalizations of the templates for the direct and inverse components determined from the fitting procedure, in units of photons ${\rm cm}^{-2} \; s^{-1}$ at Earth. One should note that the ratio between the contributions of the direct and inverse components is given by $a[0]_{\rm dir}/a[0]_{\rm inv}$ and so is determined by the fitting procedure.

Equation~\ref{sum-inv-dir} can be written as:
\begin{eqnarray}
\frac{{\rm d}{\tilde \phi}_{\rm tot}(E)}{{\rm d}E}= \frac{a[0]_{\rm dir}}{I_{\rm dir}}\left[\frac{{\rm d} \phi_{\rm dir}(E)}{{\rm d}E} + \frac{{\rm d}\phi'_{\rm inv}(E)}{{\rm d}E}\right] \; ,
\end{eqnarray}
\noindent
where
\begin{eqnarray}
\frac{{\rm d}\phi'_{\rm inv}(E)}{{\rm d}E}= \left(\frac{a[0]_{\rm inv}I_{\rm dir}}{a[0]_{\rm dir}I_{\rm inv}}\right)\frac{{\rm d}\phi_{\rm inv}(E)}{{\rm d}E} \; .
\end{eqnarray}
\noindent
This corresponds to the contribution to the total model spectrum one would obtain from a single template built by combining the direct and inverse spectra given in Equations~\ref{dir-spec} and \ref{inv-spec} with the relative abundances of the primary accelerated heavy ions multiplied by a common factor, {\it i.e.}, $w_i \rightarrow w'_i= w_{\rm fac}w_i$, where
\begin{eqnarray}
w_{\rm fac}=\left(\frac{a[0]_{\rm inv}I_{\rm dir}}{a[0]_{\rm dir}I_{\rm inv}}\right) \; .
\label{factor}
\end{eqnarray}

The total yield of this single template is given by
\begin{eqnarray}
I'_{\rm tot} &=& I_{\rm dir} + I'_{\rm inv}=\int_{E_{\rm min}^{\rm phot}}^{E_{\rm max}^{\rm phot}}\left[\frac{{\rm d}\phi_{\rm dir}(E)}{{\rm d}E}+ \frac{{\rm d}\phi'_{\rm inv}(E)}{{\rm d}E}\right] {\rm d}E \nonumber \\
&=&I_{\rm dir}+\frac{a[0]_{\rm inv}}{a[0]_{\rm dir}}I_{\rm dir}= \left(1+\frac{a[0]_{\rm inv}}{a[0]_{\rm dir}}\right)I_{\rm dir}\;.
\end{eqnarray}

It follows from the above discussion that the number of primary accelerated protons at the Sun with energies $> 30 \; {\rm MeV}$ is given by:
\begin{equation}
N_p (> 30 \; {\rm MeV})=a[0]_{\rm dir}\frac{4\pi R^2}{I'_{\rm tot}}\Delta t=a[0]_{\rm dir}\frac{4\pi R^2}{\left(1+\frac{a[0]_{\rm inv}}{a[0]_{\rm dir}}\right)I_{\rm dir}}\Delta t\; ,
\end{equation}
\noindent
where $R=1\;{\rm AU}$ and $\Delta t$ is the time interval over which the observed spectrum is integrated. The corresponding numbers of primary accelerated $\alpha$-particles and heavy ions of species $i$ are respectively given by $N_{\alpha} = w_{\alpha}N_p$ and $N_i= w'_i N_p$.

The GOES M2-class solar flare of 2010 June 12 (SOL2010-06-12T00:57) was observed by the two instruments of the {\it Fermi} satellite, the {\it Gamma-ray Burst Monitor} (GBM) and the {\it Large Area Telescope} (LAT), in the NOAA active region 11081 at the heliographic coordinates  N23W43  \cite[]{Ackermann2012}. In our analysis of this event we use the Continuous Spectroscopy (CSPEC) data from the GBM {\it bismuth germanate} (BGO) detector \cite[]{Meegan2009} and the LAT Low Energy (LLE) data from LAT \cite[]{Pelassa2010}.

Using OSPEX, we fit the GBM background-subtracted spectrum in the energy range from $0.3$ to $10\;{\rm MeV}$, accumulated between 00:55:40 and 00:58:50 UT\footnote{One should note that in \cite{Ackermann2012} they fit the background-subtracted GBM and LAT spectra integrated over a 50 s time interval (00:55:40 - 00:56:30 UT), but for the delayed neutron capture and annihilation lines they fit the spectra integrated over a 250 s time interval (00:55:40 - 00:59:50 UT). In order to account for the delay of the neutron capture and annihilation lines, we choose to fit the spectra integrated over a single larger time interval of 190 s (00:55:40 - 00:58:50 UT) for all spectral components.} and the LAT background-subtracted spectrum in the energy range from $30$ to $300\;{\rm MeV}$ accumulated in the same time interval. We fit these spectra with four components: the two FLUKA templates for the direct and inverse components of the full $\gamma$-ray spectrum and, following the analysis by \cite{Ackermann2012}, a single power-law function (1pow) and a power-law function multiplied by an exponential (1pow-exp). The last two components account for the bremsstrahlung of primary accelerated electrons and are both provided as standard fitting components by OSPEX.

Because OSPEX does not currently have the capability to perform a joint simultaneous analysis of data obtained with more than one instrument, we make use of an iterative procedure, fixing some of the parameters using data from one instrument and then refining values of others using data from the other, repeating this to iterate to a single solution using both datasets, characterised by a minimum value of the reduced chi-squared statistic. We make no claim of uniqueness for the data fits obtained below; their main point is to show that FLUKA simulations can provide photon spectra that give a statistically acceptable fit to data across the whole of the $\gamma$-ray range, for reasonable values of fast ion numbers and energy distribution.

\clearpage
\newpage
\pagebreak

We start by choosing the parameters of the function 1pow-exp at values that provide a contribution to the flux similar to that obtained by \cite{Ackermann2012}. This first step is necessary because the 1pow-exp component competes in the fitting procedure with the FLUKA template for the component due to the inverse reactions so, if one allows OSPEX to fit all four of the spectral components at once, it attributes too much of what should be the flux from inverse reactions to the 1pow-exp component instead\footnote{Similar issues are mentioned by \cite{Ackermann2012} although they use a single nuclear template.}. The parameters of the 1pow-exp component are then held fixed throughout the rest of the iterative procedure which consists of the following steps:

i) The GBM spectrum is fitted by varying the normalization parameters of the two FLUKA templates and the parameters of the function 1pow, while keeping fixed the parameters of the function 1pow-exp.

ii) The LAT spectrum is fitted by varying the normalization parameters of the two FLUKA templates, now fixing the parameters of the functions 1pow and 1pow-exp. We allow the normalisation parameters of the two FLUKA templates to vary only within a restricted range around the values found in the previous step because the relative importance of direct and inverse components is much more strongly constrained by GBM data than by LAT.

iii) The GBM spectrum is fitted again by varying the parameters of the function 1pow, while keeping fixed the normalization parameters of the two FLUKA templates obtained in step (ii) and the parameters of the function 1pow-exp.

iv) Steps (ii) and (iii) are repeated iteratively until the best-fit to the combined GBM/LAT spectrum is obtained.

\begin{figure}[ht]
 \centering
               \includegraphics[width=1\textwidth,clip=]{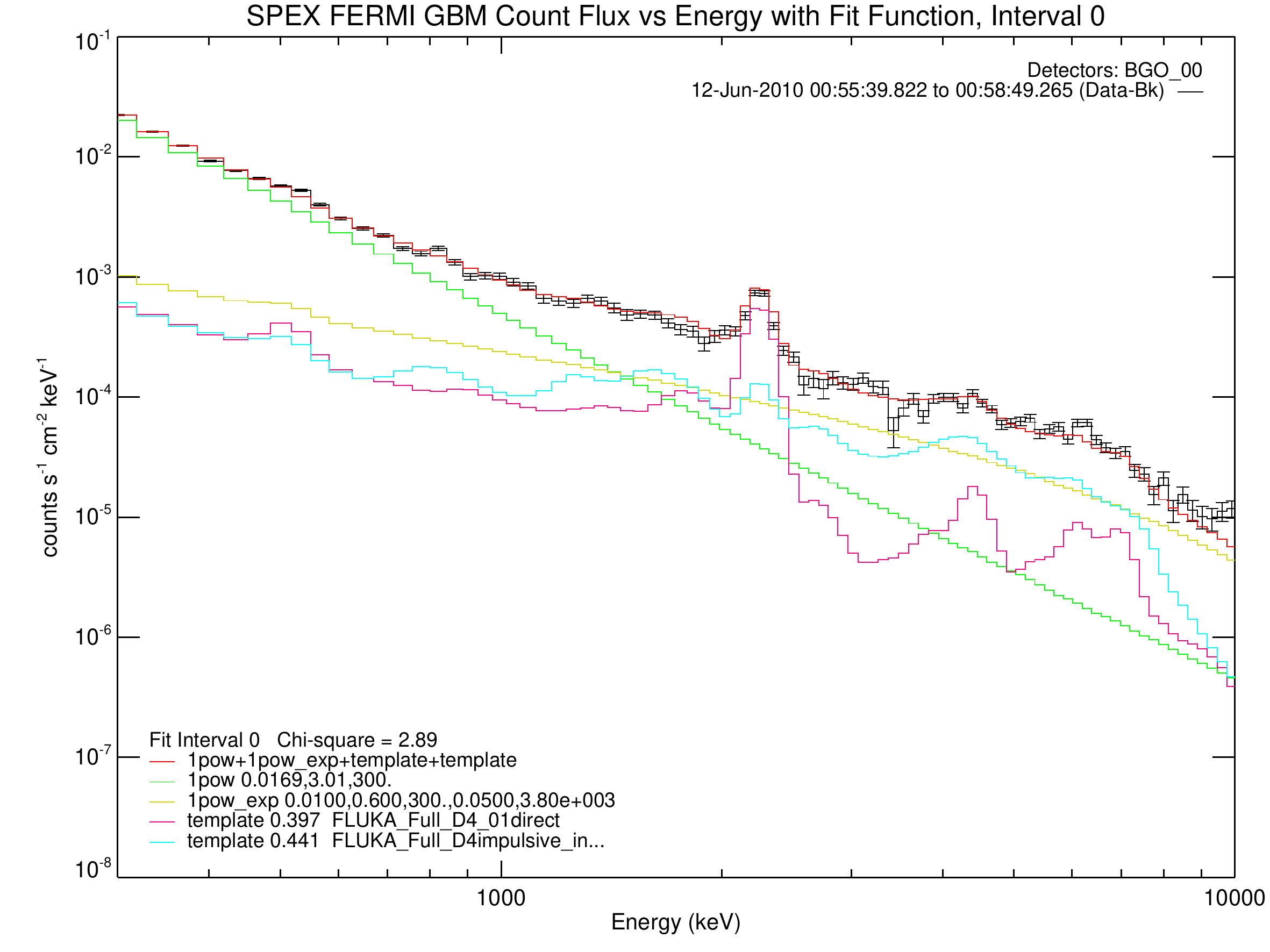} \quad
               \includegraphics[width=1\textwidth,clip=]{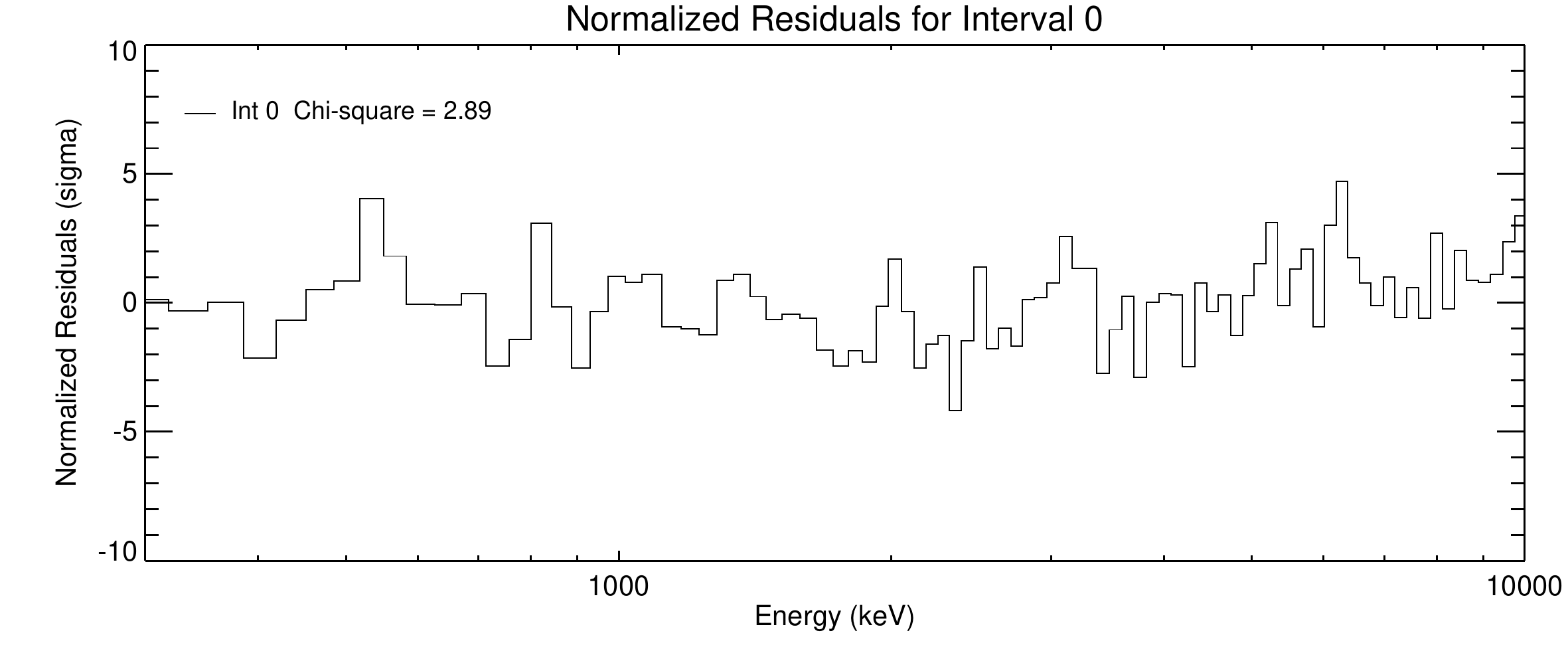}
   \caption{Fit to the background-subtracted GBM spectrum of the 2010 June 12 solar flare obtained with the FLUKA templates for the full $\gamma$-ray spectrum components. Top panel: count spectrum; Bottom panel: normalized residuals.}
  \label{F-12jun-FLUKA-Full-counts-GBM}
   \end{figure}

\begin{figure}[ht]
  \centering
               \includegraphics[width=1\textwidth,clip=]{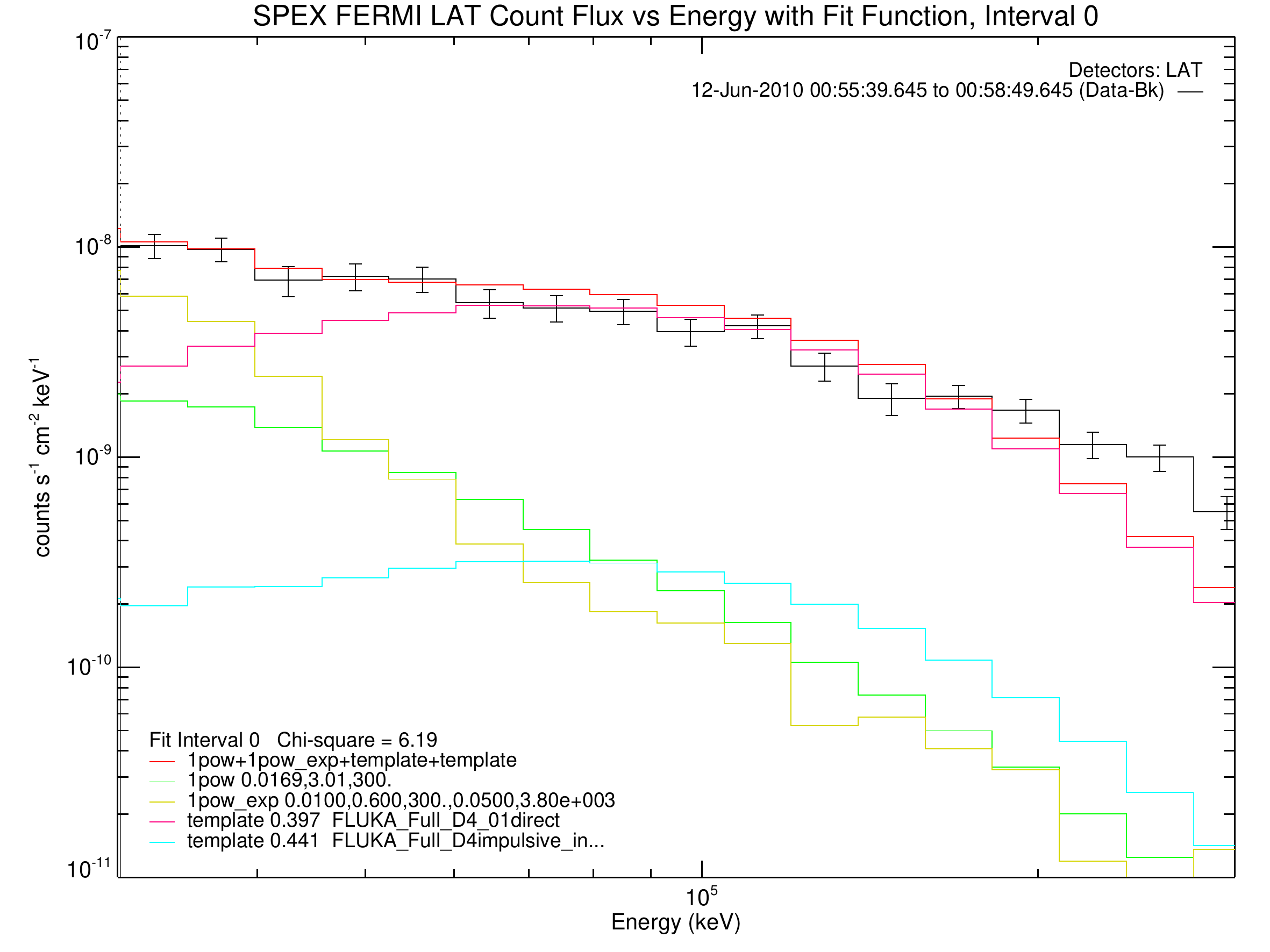} \quad
               \includegraphics[width=1\textwidth,clip=]{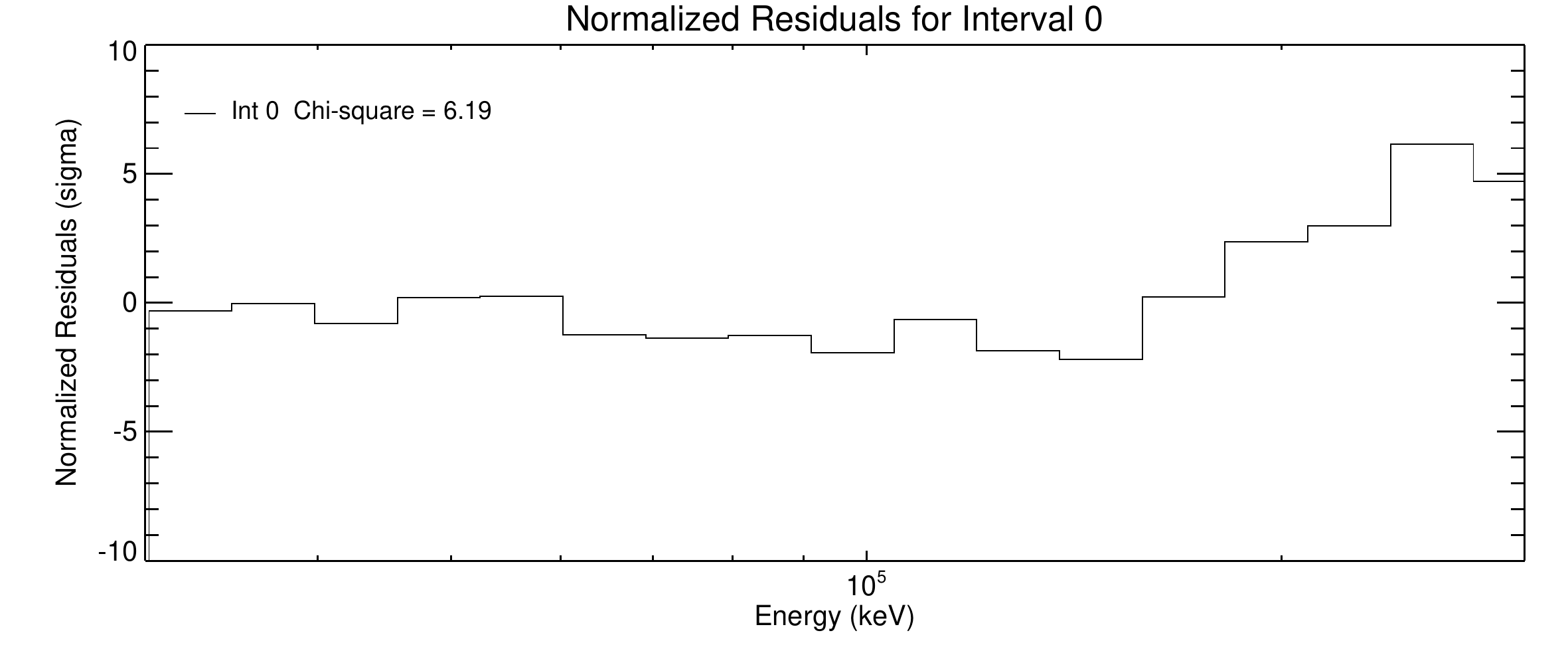}
   \caption{Fit to the background-subtracted LAT spectrum of the 2010 June 12 solar flare obtained with the FLUKA templates for the full $\gamma$-ray spectrum components. Top panel: count spectrum; Bottom panel: normalized residuals.}
   \label{F-12jun-FLUKA-Full-counts-LAT}
   \end{figure}

\begin{figure}[ht]
     \centering
               \includegraphics[width=1.0\textwidth,clip=]{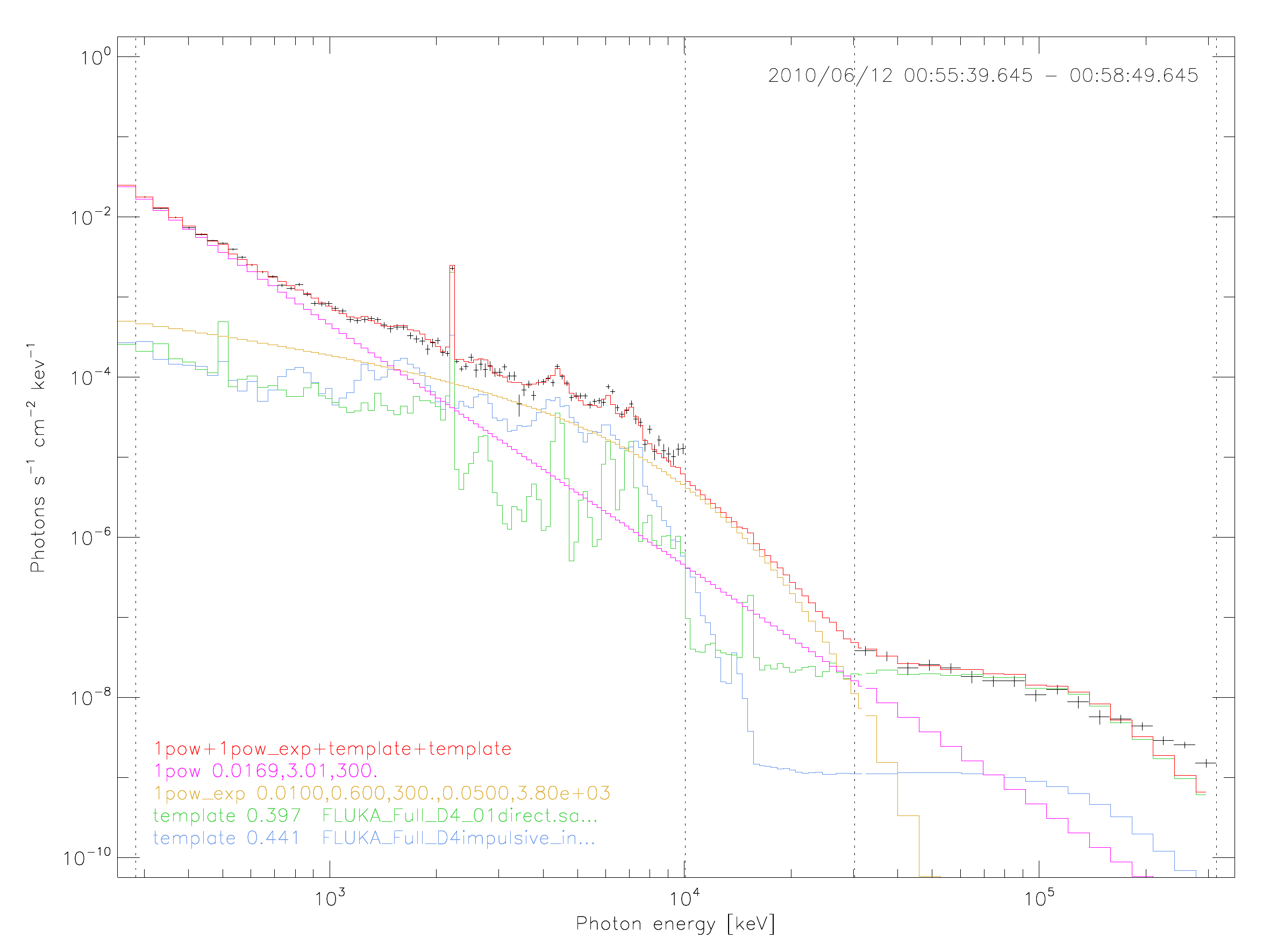}
   \caption{Best-fit to the combined GBM/LAT photon spectrum of the 2010 June 12 solar flare obtained with the FLUKA templates for the full $\gamma$-ray spectrum components. The four best-fitting components are also shown. The vertical dashed lines indicate the energy ranges used in the fits to the GBM and LAT data.}
   \label{F-12jun-FLUKA-full}
   \end{figure}

In Figures \ref{F-12jun-FLUKA-Full-counts-GBM} and \ref{F-12jun-FLUKA-Full-counts-LAT} we show the results for the fits to the GBM and LAT spectra of the 2010 June 12 solar flare obtained with the FLUKA templates for the full $\gamma$-ray spectrum components. In Figure \ref{F-12jun-FLUKA-full} we show the result for the best-fit to the combined GBM/LAT photon spectrum. The four best-fitting components are also shown. The model parameters obtained in this fit are listed in Table \ref{tab:parametros_ajust12jun_FLUKA_Full} together with their estimated uncertainties (as mentioned above the parameters of the 1pow-exp function are adopted at the outset and held fixed during the fitting procedure so we give no uncertainties for them). We note that the parameters of the 1pow component are in reasonable agreement with those found by \cite{Ackermann2012}, allowing for the difference in accumulation times. Also shown in the table are the values for the ratio between the contributions of the direct and inverse components, $a[0]_{\rm dir}/a[0]_{\rm inv}$, the common factor that multiplies the relative abundances of the primary accelerated heavy ions, $w_{\rm fac}$, and the number of primary accelerated protons, $N_p(> 30 \; {\rm MeV})$, obtained from the fit.

\clearpage
\newpage
\pagebreak

\clearpage
\newpage
\pagebreak

\clearpage
\newpage
\pagebreak

\begin{table}[ht]
\caption{Parameters of the best-fit to the combined GBM/LAT spectrum of the 2010 June 12 solar flare obtained with the FLUKA templates for full $\gamma$-ray spectrum components.}
\begin{tabular}{l c}
\hline
  Best-fit parameter & Value   \\
\hline
    normalization of the function 1pow at $300 \; {\rm keV}$                      & $0.0169 \pm 0.0003\; {\rm cm^{-2}\; keV^{-1}\; s^{-1}}$  \\
    spectral index $\delta_{e}^{\rm 1pow}$ of the function 1pow         & $- ~3.01 \pm 0.03$                 \\
    normalization of the function 1pow-exp at $300 \; {\rm keV}$                & $0.01 \; {\rm cm^{-2}\; keV^{-1}\; s^{-1}}$  \\
    spectral index $\delta_{e}^{\rm exp}$ of the function 1pow-exp        & $- ~0.600$                 \\
    normalization of the exp. of function 1pow-exp                             & $0.05 \; {\rm cm^{-2}\; keV^{-1}\; s^{-1}}$ \\
    pseudo-temp. $T_{e}^{\rm exp}$ of the function 1pow-exp                                 & $3800 \; {\rm keV}$               \\
    normalization $a[0]_{\rm dir}$ of the direct comp. template     & $0.397 \pm 0.018\; {\rm cm^{-2}\; s^{-1}}$  \\
    normalization $a[0]_{\rm inv}$ of the inverse comp. template     & $0.441 \pm 0.018\; {\rm cm^{-2}\; s^{-1}}$  \\
\hline
Calculated parameter & Value \\
\hline
    $a[0]_{\rm dir}/a[0]_{\rm inv}$                           & $0.90$      \\
    $w_{\rm fac}$                                             & $0.12$      \\
    $N_p(> 30 \; {\rm MeV})$                                  & $9.22 \times 10^{31}$     \\
\hline
\end{tabular}
\label{tab:parametros_ajust12jun_FLUKA_Full}
\end{table}

\clearpage
\newpage
\pagebreak

It is useful to comment separately on the quality of the fit in the GBM and LAT energy ranges. For the GBM spectrum in the energy range from $300 \; {\rm keV}$ to $10 \; {\rm MeV}$ the fit is very reasonable ($\chi_{r}^{2}= 2.89$), particularly for the structures corresponding to the $\gamma$-ray lines produced by the de-excitation of $^{12}$C and $^{16}$O nuclei. As shown in Figure \ref{F-12jun-FLUKA-Full-counts-GBM}, the normalization $a[0]_{\rm dir}$ of the FLUKA template for the component due to the direct reactions is fixed primarily by the fit to the structure corresponding to the $2.223\;{\rm MeV}$ neutron-capture line while the normalization $a[0]_{\rm inv}$ of the FLUKA template for the component due to the inverse reactions is determined mainly by the fit to the structures corresponding to the $^{12}$C and $^{16}$O lines. As discussed above (Section~\ref{S-Lines}) our FLUKA template for the component due to the direct reactions exhibits less well-defined structures for the narrow lines between $\approx 1$ to $2 \; {\rm MeV}$ range but the GBM data also do not show well-defined narrow lines in this energy range so there is little adverse impact on the quality of the fit. The predicted $2.223 \; {\rm MeV}$ line appears broader than expected but this detail, also found in \cite{Ackermann2012}, seems to reflect details of the provided GBM instrument response matrix.

The result obtained for the fit to the LAT spectrum in the photon energy range from $30$ to $300 \; {\rm MeV}$ is only satisfactory ($\chi_{r}^{2}= 6.19$). As shown in Figure \ref{F-12jun-FLUKA-Full-counts-LAT}, the fit is dominated by the component due to the direct reactions. For energies below $\approx 200 \; {\rm MeV}$ the fit is very good. For energies above $\approx 200 \; {\rm MeV}$, on the other hand, the component due to the direct reactions is weaker than that observed in the data, resulting in more significant residuals. The FLUKA templates used in these fits assume a primary ion distribution extending only to $1 \; {\rm GeV/nucleon}$. Templates constructed assuming a greater maximum ion energy would certainly fit the LAT data better at higher energies.

Since the components due to the direct and inverse reactions are strongly constrained by the fit to the GBM spectrum in the range from $300 \; {\rm keV}$ to $10 \; {\rm MeV}$, only two iterations are necessary to obtain the best fit of the combined GBM/LAT spectrum. A full analysis of these data would search for a best fit across several templates calculated using different values of ion energy spectral index $\delta$. The exercise carried out here, with a fixed value $\delta = 4$, nonetheless shows that the FLUKA simulations can give a description consistent with GBM and LAT data, across the photon energy range from $300 \; {\rm keV}$ to $300 \; {\rm MeV}$. Although the strength of the conclusion is limited by the GBM's spectral resolution, it is remarkable that a single ion spectral index can achieve an acceptable fit to the data across this wide energy range.

%%%%%%%%%%%%%%%%%%%%%%%%%%%%%%%%%%%%%%%%%%%%%%%%%%

\section{Summary and Concluding Remarks}
\label{S-summary}

We presented simulations of the $\gamma$-ray spectrum of solar flares using the Monte Carlo code FLUKA. The nuclear de-excitation $\gamma$-ray line spectra obtained in our FLUKA simulations are in reasonable agreement with the more detailed spectra generated with the RMK code, currently the main tool for analysis of the $\gamma$-ray spectrum of solar flares. Moreover, our FLUKA simulations of the full $\gamma$-ray spectrum are self-consistent, in that the main spectral components in this range (the nuclear de-excitation lines, the $511 \; {\rm keV}$ positron annihilation line -- excluding, for now, the contribution from the decay of radioactive positron-emitter daughter nuclei -- the $2.223 \; {\rm MeV}$ neutron-capture line, and the pion decay continuum) are generated from the same distribution of primary accelerated ions. The transport of accelerated ions and photons through a semi-empirical model of the ambient solar atmosphere is taken into account, along with a number of relevant nuclear processes. To the best of our knowledge, this is the first attempt to fit $\gamma$-ray spectra of solar flares using a self-consistent model of the several spectral features in the energy range from $\approx 100$s ${\rm keV}$ to $\approx 100$s ${\rm MeV}$. Previous analysis employed independent functions or templates for the nuclear de-excitation, neutron capture and annihilation lines, and the pion decay continuum. FLUKA's usefulness for this purpose will be further enhanced when future work includes the contribution of radioactive daughter nuclei to the $511 \; {\rm keV}$ positron annihilation line.

FLUKA can be a valuable tool for the study and interpretation of solar flare $\gamma$-ray spectra. Using data from {\it Fermi} (GBM and LAT) for the 2010 June 12 solar flare \cite[]{Ackermann2012}, we show that FLUKA gives a statistically acceptable fit to the $\gamma$-ray spectrum (as measured by the reduced chi-square), implying reasonable estimates of ion numbers. It is important to emphasize that in this paper we do not intend to perform a detailed analysis of solar flare $\gamma$-ray data. However we have taken some care in Section~\ref{S-fitting} to show how the spectral components appear, separately and in combination, when compared with data.

FLUKA's treatment of nuclear collisions, excitation and de-excitation is sufficiently complete to capture all the main features in the photon spectrum ({\it cf.} the discussion of GEANT4's de-excitation line capabilities in \cite{Tang2010}). In order to completely exploit these capabilities it will be necessary to improve both the statistics and the energy resolution of the photon spectra generated in the FLUKA simulations, a task we will pursue in a future work.

Our results were obtained using the standard FLUKA USRBDX detector that monitors the flux of particles crossing a defined surface. The regularly spaced bins used by USRBDX, combined with the comparatively low cross-sections for some of the de-excitation lines resulted in a less detailed photon spectrum than we might have desired below $\approx 2 \; {\rm MeV}$. {\it Via} some of the user-editable routines, it is possible to implement {\it e.g.} nonuniform bin sizes to improve the quality of the spectrum around the strongest narrow lines.

Since FLUKA evaluates the transport of both particles and photons, it is possible to investigate the directivity of the $\gamma$-ray emission, especially comparing the $2.223 \; {\rm MeV}$ neutron capture line and nuclear de-excitation lines. The first is expected to be generated deep in the photosphere and susceptible to strong limb-darkening effect \cite[]{Hua1987}, while the latter are expected to originate in the chromosphere, without limb-darkening effects. The versatility of FLUKA opens the possibility to revisit observations of $\gamma$-ray flares from past decades to reveal more information about the elusive presence of accelerated ions during solar flares, and to predict {\it e.g.} $\gamma$-ray height structure that could be studied with future instruments with imaging capability.

Each simulated spectrum shown here typically required many days' processing time on a high-end laptop computer. We relied on standard FLUKA detectors, particularly USRBDX, to obtain these spectra which are specific to the assumed energy and angular distribution of the primary ions and the chemical abundances of the target and accelerated ions. Many projects, for instance a comprehensive exploration of the parameter space, would be computationally challenging or even impossible with this approach. An alternative approach, however, would start from a well-defined set of primary ion energies and directions and store the details of all the photons crossing the boundary between lower atmosphere and corona, together with the details of the primary particle that gave rise to them. Emergent spectra for different injected ion distributions could then be synthesised by applying weights determined by the energies and directions of the primary ions to the photons obtained via full FLUKA simulations. Variations in source or target abundances could be accommodated similarly. Calculating a spectrum via application of these weights would involve only modest computational effort, likely light enough to enable synthesis of spectra on-the-fly for fitting, systematic exploration of the primary ion parameter space, {\it etc}. The major computational effort would be in obtaining the basic photon spectra with enough statistical reliability in the first place but this would only need to be done once. The present work has demonstrated that FLUKA simulations can be a valuable tool for interpreting flare $\gamma$-ray data; such a project, to be pursued elsewhere, offers a practical approach to making them routinely useful.

\begin{acks}
The referees' many comments and questions helped greatly in improving the paper. We thank the Royal Society Newton Fund for supporting our UK-Brazil collaboration through project NI140209 \emph{The THz wavelength range as a window on extremes of solar flare particle acceleration}. We also thank the financial support from FAPESP under grants 2009/18386-7 and 2017/13282-5. PJAS acknowledges support from the University of Glasgow's Lord Kelvin Adam Smith Leadership Fellowship. ALM and PJAS acknowledge relevant and helpful discussions with members of the ISSI International Team on \emph{Energetic Ions: The Elusive Component of Solar Flares} and with participants in the Lorentz Center Workshop on \emph{Solar Sources of GeV Gamma-rays}, 26 Feb - 2 Mar 2018. DST acknowledges support from CAPES and Instituto Presbiteriano Mackenzie. CGGC research is partially supported from CNPq (Grant 305203/2016-9). We thank the FLUKA team who provide and support the code. Finally, we thank R. J. Murphy for kindly supplying the copy of the RMK code for calculating nuclear de-excitation $\gamma$-ray line spectra we used in this work.
\end{acks}

\section*{Disclosure of Potential Conflicts of Interest}

The authors declare that they have no conflicts of interest.

\newpage

\appendix

FLUKA simulations are set up by means of an input ASCII file built by the user which consists of a sequence of command lines, often called ``cards'' for historical reasons \cite[]{Ferrari2011}. In general, a typical FLUKA input file contains cards through which the following simulation elements are defined:

\begin{itemize}
\item Particle source properties: characteristics of the beam of primary particles such as particle type, energy and angular distribution, starting position);
\item Geometry: solid bodies and surfaces combined by boolean operations to implement the complete partition of the space of interest into regions;
\item Materials: single-elements or compounds (either pre-defined or user-defined) which are assigned to the geometry regions;
\item Detectors: estimators used to score physical quantities of interest;
\item Settings: parameters, conditions and general directives which determine how calculations are performed, {\it e.g.} specification of production and transport thresholds, biasing schemes and physical effects to be included.
\end{itemize}

FLUKA can accurately simulate the transport and the interactions of about $60$ different types of particles, including electrons and muons with energies from $1 \; {\rm keV}$ to $1000 \; {\rm TeV}$, photons with energies from $100 \; {\rm eV}$ to $10000 \; {\rm TeV}$, hadrons with energies from $1 \; {\rm keV}$ to $10000 \; {\rm TeV}$, and all corresponding antiparticles, as well as neutrinos, low-energy neutrons (below $20 \; {\rm MeV}$) down to thermal energies and heavy ions with energies up to $10000 \; {\rm TeV/nucleon}$. Interactions are implemented by robust and updated physics-models, continuously benchmarked and optimized against experimental data at single interaction level. The models are fully integrated, such as to provide a consistent treatment of all physical processes to the same level of accuracy.

The models used in FLUKA to implement the transport and interactions of electromagnetic particles and muons cover a comprehensive range of processes, including bremsstrahlung, Compton and Rayleigh scattering, pair production, positron annihilation in flight and at rest, Bhabha and Moller scattering, photoelectric effect, photonuclear interactions and photomuon production. Transport and ionization energy losses are performed through approaches shared by all charged particles (electrons, positrons, muons, charged hadrons and heavy ions). Transport is implemented through an original algorithm \cite[]{Ferrari1992} based on the Moliere theory of multiple scattering improved by \cite{Bethe1953}, supplemented by a single scattering algorithm based on the Rutherford formula. The treatment of ionization energy losses is based on the Bethe-Bloch theory \cite[]{Bethe1934}, supplemented with ionization potentials and density effect parameters taken according to the compilation of \cite{Sternheimer1984} and with shell corrections derived from the parametrized formula by \cite{Ziegler1977}. Ionization fluctuations are implemented through an original approach (alternative to the standard ones based on the Landau and Vavilov theory) which makes use of general statistical properties of the cumulants of a distribution \cite[]{Fasso1997}. The approach allows to combine explicit $\delta$-ray production and transport with ionization fluctuations and also to include corrections for spin-relativistic effects and distant collisions.

The treatment of hadronic interactions in FLUKA is based on a microscopic approach which makes use of several models properly tailored to the different energy ranges \cite[]{Ferrari1998,Battistoni2015}. At energies below $5 \; {\rm GeV}$, inelastic hadron-hadron interactions are implemented through the isobar model based on the resonance production and decay of particles, while elastic and charge exchange interactions are implemented through phase-shift analysis and eikonal approximation\footnote{At energies below the pion production threshold ($\approx 290 \; {\rm MeV}$ for nucleon-nucleon interactions and $\approx 170 \; {\rm MeV}$ for pion-nucleon interactions), only elastic and charge exchange scattering processes take place.}. At energies in the range from $5$ to $20 \; {\rm TeV}$, inelastic hadron-hadron interactions are implemented through a modified version of the {\it Dual Parton Model} (DPM) \cite[]{Capella1994} coupled to a hadronization scheme. Inelastic hadron-nucleus interactions at energies from reaction threshold up to  $20 \; {\rm TeV}$ are implemented through the FLUKA model called PEANUT (Pre-Equilibrium Approach to Nuclear Thermalization) \cite[]{Ferrari1994,Fasso1994,Ferrari1998,Battistoni2006}.

The PEANUT model describes the interactions of hadrons with nuclei as a sequence of the following steps:

\begin{itemize}
\item Glauber-Gribov multiple scattering cascade with formation zone;
\item Generalized intra-nuclear cascade;
\item Pre-equilibrium emission stage (exciton-based);
\item Equilibrium stage: evaporation, fission, Fermi break-up, $\gamma$ de-excitation.
\end{itemize}
\noindent
The $\gamma$ de-excitation process is particularly important for our simulations of $\gamma$-ray spectra from solar flares. After the evaporation process, the residual excitation energy of nuclei is dissipated through the emission of nuclear de-excitation $\gamma$-ray photons. At high excitation energies the cascade of $\gamma$-ray transitions is implemented through a statistical model by assuming a continuous density of nuclear levels, while at low excitation energies (below an arbitrary threshold) a database of tabulated discrete experimental nuclear levels is used \cite[]{Ferrari1996}.

The extension from hadron-nucleus to nucleus-nucleus interactions is implemented by modified versions of three external event generators linked to FLUKA, namely the Boltzmann Master Equation (BME) model \cite[]{Cerutti2006} at energies below $0.125 \; {\rm GeV/nucleon}$, the Relativistic Quantum Molecular Dynamics (RQMD-2.4) model \cite[]{Sorge1989,Andersen2004} at energies in the range from $0.125$ to $5 \; {\rm GeV/nucleon}$, and the Dual Parton Model and Jets (DPMJET-III) model \cite[]{Roesler2001} at energies above $5 \; {\rm GeV/nucleon}$. All three external event generators are interfaced with the PEANUT model in order to handle the equilibrium stage processes.

We end this section with a brief description of the treatment given in FLUKA to the low-energy neutrons ({\it i.e.}, those with energies in the range from thermal to $20 \; {\rm MeV}$), which is also relevant for our simulations of $\gamma$-ray spectra from solar flares since it provides the emission of the $2.223 \; {\rm MeV}$ neutron capture line. Transport and interactions of low-energy neutrons are implemented in FLUKA through a multigroup algorithm \cite[]{Ferrari2011,Battistoni2015}. The scattering matrices for the reaction channels are calculated from neutron cross section data provided by a dedicated library in which the energy range of interest is divided into $260$ neutron energy groups ($31$ of which are thermal) and $42$ gamma energy groups. Elastic and inelastic reactions are simulated by group-to-group transfer probabilities. In particular, the production of $\gamma$-ray photons by low-energy neutrons (including the $2.223 \; {\rm MeV}$ neutron capture line) is treated by evaluating a downscattering matrix which provides the probability for a neutron in a given group to generate a photon in each of the gamma groups\footnote{One should note that the multigroup algorithm is used only for gamma generation. The transport of $\gamma$-ray photons produced by low-energy neutrons is performed through the same modules used in FLUKA for the transport of photons produced by other processes.}.

\bibliographystyle{spr-mp-sola}

%\bibliography{sola_bibliography}

%\end{article}

%\end{document}

\end{article}

\end{document}